%% file: main.tex
\documentclass[twocolumn]{aastex63}

\input preamble.tex
\shorttitle{GDIGS Discrete Sources}
\shortauthors{Linville et al.}
\begin{document}

\title{The GBT Diffuse Ionized Gas Survey (GDIGS): Discrete Sources}

\author[0000-0002-4727-7619]{Dylan~J.~Linville}
\affiliation{Department of Physics and Astronomy, West Virginia University, Morgantown, WV 26506, USA}
\affiliation{Center for Gravitational Waves and Cosmology, West Virginia University, Chestnut Ridge Research Building, Morgantown, WV 26505, USA}

\author[0000-0001-8061-216X]{Matteo~Luisi}
\affiliation{Center for Gravitational Waves and Cosmology, West Virginia University, Chestnut Ridge Research Building, Morgantown, WV 26505, USA}
\affiliation{Department of Physics, Westminster College, New Wilmington, PA 16172, USA}

\author[0000-0001-8800-1793]{L.~D.~Anderson}
\affiliation{Department of Physics and Astronomy, West Virginia University, Morgantown, WV 26506, USA}
\affiliation{Center for Gravitational Waves and Cosmology, West Virginia University, Chestnut Ridge Research Building, Morgantown, WV 26505, USA}
\affiliation{Adjunct Astronomer at the Green Bank Observatory, P.O. Box 2, Green Bank, WV 24944, USA}

\author[0000-0002-1311-8839]{Bin Liu}
\affil{National Astronomical Observatories, Chinese Academy of Sciences, Beijing 100012, China}
\affiliation{CAS Key Laboratory of FAST, NAOC, Chinese Academy of Sciences, China}

\author[0000-0003-4866-460X]{T.~M.~Bania}
\affiliation{Institute for Astrophysical Research, Astronomy Department, Boston University, 725 Commonwealth Ave., Boston, MA 02215, USA}

\author[0000-0002-2465-7803]{Dana.~S.~Balser}
\affiliation{National Radio Astronomy Observatory, 520 Edgemont Road, Charlottesville, VA 22903, USA}

\author[0000-0003-0640-7787]{Trey~V.~Wenger}
\affiliation{Dominion Radio Astrophysical Observatory, Herzberg Astronomy and Astrophysics Research Centre, National Research Council, P.O. Box 248, Penticton, BC V2A 6J9, Canada}
\affiliation{ NSF Astronomy \& Astrophysics Postdoctoral Fellow, Department of Astronomy, University of Wisconsin-Madison, Madison, WI 53706, USA}

\author[0000-0002-9947-6396]{L.~M.~Haffner}
\affiliation{Department of Physical Sciences, Embry-Riddle Aeronautical University, Daytona Beach FL 32114, USA}

\author[0000-0002-3758-2492]{J.~L.~Mascoop}
\affiliation{Department of Physics and Astronomy, West Virginia University, Morgantown, WV 26506, USA}
\affiliation{Center for Gravitational Waves and Cosmology, West Virginia University, Chestnut Ridge Research Building, Morgantown, WV 26505, USA}

\correspondingauthor{Dylan~J.~Linville}
\email{djl0023@mix.wvu.edu}

\begin{abstract}
The Green Bank Telescope (GBT) Diffuse Ionized Gas Survey (GDIGS) traces ionized gas in the Galactic midplane by observing radio recombination line (RRL) emission from 4--8\,\ghz. The nominal survey zone is $32.3\degree>\ell >-5\degree$, $\absb<0.5\degree$.
Here, we analyze GDIGS \hna\ ionized gas emission toward discrete sources. 
Using GDIGS data, we identify the velocity of 35 \hii regions that have multiple detected RRL velocity components.  
We identify and characterize RRL emission from 88 \hii regions that previously lacked measured ionized gas velocities.
We also identify and characterize RRL emission from eight locations that appear to be previously-unidentified \hii regions and 30 locations of RRL emission that do not appear to be \hii\ regions based on their lack of mid-infrared emission. This latter group may be a compact component of the Galactic Diffuse Ionized Gas (DIG).
There are an additional 10 discrete sources that have anomalously high RRL velocities for their locations in the Galactic plane. We compare these objects' RRL data to \cor, \hi and mid-infrared data, and find that these sources do not have the expected 24\,\micron\ emission characteristic of \hii regions. 
Based on this comparison we do not think these objects are \hii regions, but we are unable to classify them as a known type of object.
\end{abstract}


\section{Introduction}

Ionized gas in the plane of the Milky Way can be decomposed into discrete clumps, particularly \hii regions, and a diffuse component known as the ``Diffuse Ionized Gas'' (DIG).
\hii\ regions are zones of ionized gas that form around spectral type O and B stars, where the star's immense energy output ionizes the surrounding medium. These massive stars only live $\sim\!10\,\myr$. \hii\ regions are therefore useful indicators of active massive star formation.  
In contrast to \hii\ regions, the DIG is spread throughout the Galactic disk, has a lower electron density near $n_e\!\sim\!10^{-2}\percc$ \citep{Lyne85} compared to densities of $n_e\!\sim\!10^{2}\percc$ or greater found in \hii\ regions \citep{Lockman75}, and is not typically associated with specific stars.

We can study Galactic ionized gas using transitions of excited hydrogen atoms. Recombination lines are produced after electrons and ions recombine into an excited state. For nearby plasma, H$\alpha$ line observations such as those made by the Wisconsin H$\alpha$ Mapper (WHAM) survey \citep{reynolds98, haffner03} provides sensitive observations that can determine gas properties.
The H$\alpha$ line, although bright, is attenuated by dust and so is of limited use for distant \hii\ regions. 
Radio recombination lines (RRLs) are less affected by extinction compared to H$\alpha$ and so are the preferred tracer for more distant \hii\ regions and the DIG.

Previous targeted RRL surveys have been instrumental in assembling the catalog of known \hii regions. In particular, the \hii Region Discovery Survey \citep[HRDS;][]{anderson2011} and Southern \hii Region Discovery Survey \citep[SHRDS;][]{wenger21} have more than doubled the number of known \hii regions \citep{anderson18}. An ionized gas spectroscopic detection determines the source's velocity relative to the local standard of rest (LSR).  Such pointed observations, however, have an inherent weakness: they are by nature biased toward sources meeting the criteria for inclusion in the study.

Large-scale mapped RRL surveys allow for the study of ionized gas across the Galaxy. The Green Bank Telescope (GBT) Diffuse Ionized Gas Survey (GDIGS) is a RRL survey that was conducted with the goal of studying ionized gas in the Galactic midplane.
GDIGS offers several advantages over previous blind RRL surveys in the survey zone. Such surveys lacked the sensitivity and resolution necessary to detect many new discrete sources.  The Survey of Ionized Gas in the Galaxy, made with the Arecibo Telescope (SIGGMA), an RRL survey near $1.4\,\ghz$, has a sensitivity of $\sim\!1$\,m\!\jyb at 5\,\kms\ spectral resolution and $\sim\!3\arcmper4$ spatial resolution \citep{liu13, liu19}.  The \hi Parkes All Sky Survey (HIPASS) RRL survey, has 6.4\,m\!\jyb\ sensitivity at 20\,\kms\ spectral resolution and $14\arcmper4$ spatial resolution, also near $1.4\,\ghz$ \citep{alves15}. As detailed subsequently in Section~\ref{sec:data}, GDIGS improves upon both of these surveys in sensitivity, spatial resolution, and spectral resolution. We can use GDIGS to partially correct for the intrinsic coverage bias of pointed surveys by locating otherwise unidentified sources within the survey zone, $32.3\degree>\ell >-5\degree$, $\absb<0.5\degree$. GD
IGS is a blind survey, and is consequently sensitive to populations that targeted surveys miss. 
This feature of mapped surveys like GDIGS is especially relevant for studying clumps of the DIG, which the inclusion criteria of targeted surveys ignore.
Using mapped surveys to study \hii\ populations has been shown to be effective in \citet{Hou22}. \citet{Hou22} mapped the Galactic plane in RRL emission near 1\,\ghz\ over $55\degree>\ell >33\degree$, $\absb<2.0\degree$ with a sensitivity of 0.25\,m\!\jyb, at 2.2\,\kms spectral resolution and $\sim\!3^{\prime}$ spatial resolution using the Five-hundred-meter Aperture Spherical radio Telescope (FAST). This survey does not overlap with the GDIGS survey area.
With GDIGS we are therefore able to obtain a more complete census of \hii regions and assorted DIG features in the observed section of the Galactic disk than with previous targeted surveys.

Here we use GDIGS data to investigate new sources of RRL emission and to re-analyze previously known sources in ways that only a blind survey allows.  
We distinguish the emission from known \hii\ regions using version 2.4 of the WISE Catalog of Galactic \hii regions \citep[][hereafter the ``WISE Catalog'']{anderson14}.
The WISE Catalog is statistically complete for \hii regions ionized by O-type stars \citep{armentrout17}. 
One goal with this study is to improve the WISE Catalog by providing RRL detections and associated information, particularly velocity measurements, for as many sources previously lacking such observations as possible within the GDIGS survey area. We also intend to locate any previously unreported \hii\ regions in the survey area. In the course of the study we encountered numerous regions of discrete RRL emission that do not fit the previously described characteristics of \hii\ regions. We report these in Sections \ref{compact_dig} and \ref{sec:AVFs}.

%% %% %% %% %% %% %% %% %% %% %%
\section{Data \label{sec:data}}

\subsection{GDIGS Data}
GDIGS is a survey of ionized gas in the midplane of the Milky Way. The GDIGS RRL data were first published in \citet{anderson21}. A complete description of the details of data collection and processing can be found there, while a shorter summary is provided here. 

The GDIGS data were collected using the C-band receiver on the Green Bank Telescope (GBT) in total power mode.  Within the 4--8\,\ghz\ bandpass, we tuned to 15 usable hydrogen RRLs, as defined in \citet{anderson21}. The reduced data of average hydrogen RRL emission have a spatial resolution of 2\arcmper65 and $0.5\,\kms$ spectral resolution. 
The rms spectral noise per spaxel is $\sim\!10$\,mK, with sensitivity to emission measures EM\,$\gtrsim 1100$\,pc\,cm$^{-6}$. This emission measure translates to a mean electron density of $\langle n_e \rangle \gtrsim 30\,\percc$ for a 1\,pc path length, or $\langle n_e \rangle \gtrsim 1\,\percc$ for a 1\,kpc path length. Our data processing methodology is described in \citet{anderson21}.

\subsection{MIR Data} \label{sec:MIR_data}
Mid-infrared (MIR) observations are an important diagnostic tool used to determine if an ionized gas source is an \hii\ region or another class of object. We use 3.6\,\micron\ and 8.0\,\micron\ data from the Spitzer GLIMPSE survey \citep{benjamin03, churchwell09} and 24\,\micron\ data from the MIPSGAL survey \citep{carey09}. These surveys provide the best resolution and sensitivity available. \hii\ regions have strong emission at these wavelengths, particularly 8.0\,\micron\ and 24\,\micron. 

\subsection{The WISE Catalog\label{sec:wise}}
We compare the regions discussed in this paper to those of the WISE Catalog.
The WISE Catalog is the largest, most complete catalog of Galactic \hii\ regions, and contains all  8416 known \hii regions and \hii region candidates.
The catalog distinguishes \hii\ regions from other ionized gas sources by their characteristic emission.  

There are four types of sources listed in the catalog: ``known,'' ``candidate,'' ``group,'' and ``radio quiet.'' 
Known sources have detected spectroscopic ionized gas emission as well as  $\sim\!20\,\micron$ emission.  
Candidate sources have radio continuum emission coincident with their $\sim\!20\,\micron$ emission and are thus likely to be real \hii regions, but do not have a spectroscopic ionized gas detection. 
Radio quiet sources have MIR emission but lack associated radio continuum emission; this may be because the radio continuum emission is too faint, or because existing surveys do not cover the sources. 
Finally, group sources are located nearby ``Known'' \hii regions and are thought to be part of a larger complex but do not have measured spectroscopic ionized gas emission. 
The latter three of these categories--- ``candidate'', ``group'', and ``radio quiet'' sources- are suspected to be \hii\ regions but require confirmation before they can be moved to the ``known'' category.
The detection of RRL emission from WISE Catalog sources previously lacking such a detection provides both confirmation that the source is an \hii\ region and also an LSR velocity that can be used to derive a kinematic distance.
We use the presence of these tracers in our identification of new regions below.

The census of \hii regions in the WISE Catalog is lacking information for a large fraction of its 8416 entries, $\sim\!2000$ of which fall within the GDIGS survey area. 
Many objects have multiple reported ionized gas velocity components and therefore lack a unique LSR velocity assignment, which is a requirement for calculating a kinematic distance. 
There are 2361 confirmed \hii\ regions with spectroscopic gas detections. The WISE Catalog lacks spectroscopic gas detection for 72\% of its entries, including 626 ``group'' \hii regions, 1681 \hii region candidates, and 3748 radio-quiet \hii region candidates.

%%% %%% %%% %%% %%% %%% %%% %%% %%% %%% %%% %%% %%% %%%
\section{GDIGS Discrete Sources \label{sec:discoveries}}
GDIGS is a blind survey of ionized gas in the Galactic plane and is therefore a useful dataset both for 
characterizing previously-known \hii regions and for searching for other discrete sources of ionized gas. We examine GDIGS RRL emission from sources in the WISE Catalog to determine the velocity of the source and, where applicable, to confirm that the source is an \hii\ region. We also search the GDIGS survey data for new \hii\ regions that do not appear in the WISE Catalog, and locate numerous 
locations of compact RRL emission that lack MIR emission.

%% %%%% %% %%%% %% %%%% %% %%%%
\subsection{WISE Catalog Objects}
We explore detections in the GDIGS survey data of sources belonging to the WISE Catalog. 
Due to confusion within the GDIGS beam, 
we are only able to conclusively associate GDIGS emission with WISE Catalog sources
that are isolated or are angularly large. Smaller, more tightly grouped regions are likely to be indistinguishable from each other at the GDIGS spatial resolution. 

Some WISE Catalog sources are in close angular proximity with each other and share the same reported velocity or velocities. 
In these cases both objects were originally observed in the same telescope beam.  Following the methodology used to construct the WISE Catalog, we report the same measured velocities for both objects.

%%%%%%%%%%%%%%%%%%%%%%%%%%%%%%%%%%
\subsubsection{Multiple-Velocity H\,{\footnotesize II} Regions \label{sec:multivel}}
Many previously-known \hii\ regions have more than one measured RRL emission velocity component, which complicates efforts to calculate the regions' kinematic distances and other physical parameters. In the HRDS, for instance, over 30\% of all detected \hii\ regions have more than one RRL velocity \citep{anderson15b}. For such regions, one velocity component presumably comes from the \hii\ region itself, and the other component(s) from the DIG along the line of sight. We are confident in this assumption because two \hii\ regions along a single line of sight is statistically unlikely \citep{anderson15b}. With their large beams and sensitivity to extended emission, single-dish observations are especially susceptible to detecting emission from diffuse RRL emission in the same pointing. 

Many sources discussed in this section, both those newly detected in RRL emission and those previously known, have multiple observed velocity components. 
We can use GDIGS data with a method similar to the one described in \citet{Anderson15a} to determine the ``correct'' velocity for some of these sources. 
For multiple RRL velocity \hii\ regions in the GDIGS range we prepare Moment~0 maps of each measured velocity component by integrating over a velocity range equal to the full width at half maximum (FWHM) of the component and centered at the reported velocity. Each map is centered spatially on the WISE Catalog source. 
For each region we visually compare the Moment~0 maps at the multiple velocities; the Moment~0 morphology at the ``correct'' velocity should match the angular size of the source in the WISE Catalog. 
The emission at the other velocity component(s) along the same line of sight is expected to be more spatially extended. 

\begin{figure*}
    \centering
    \includegraphics[width=3.65in]{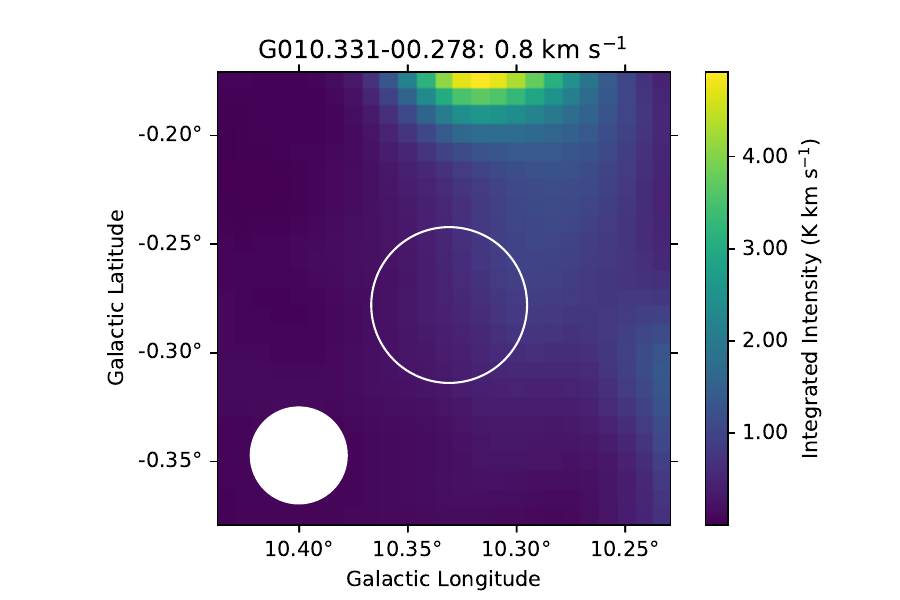}\hskip-0.3in
    \includegraphics[width=3.65in]{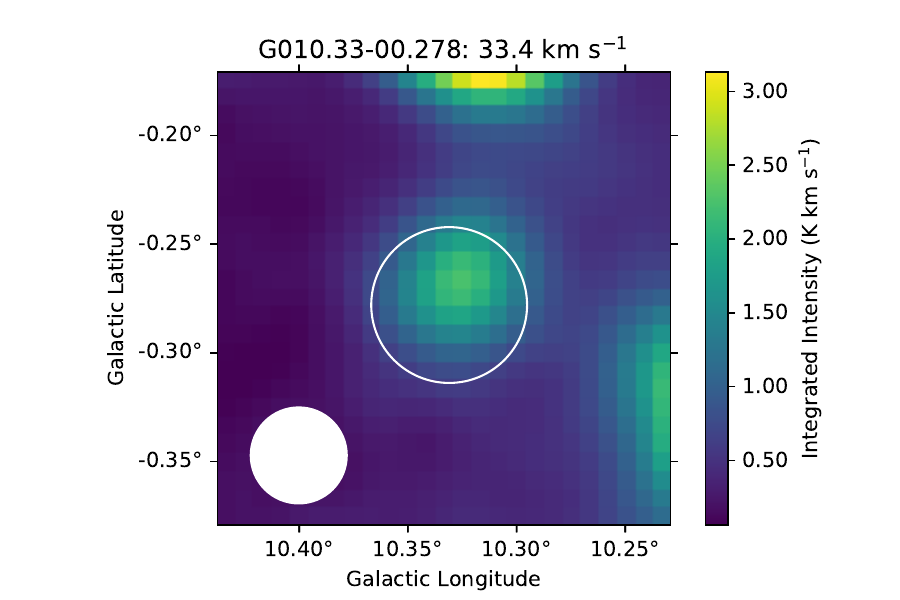}    
    \caption{
    \hna\ Moment~0 images toward source G010.331$-$0.278. Each image is integrated over a range centered at one of the two detected RRL velocities for this source and equal in width to the corresponding FWHM. The left image is 
    integrated over a FWHM of 14.1\,\kms with the range centered at 0.8\,\kms, and the right image is 
    integrated over a FWHM of 27.8\,\kms with the range centered at 33.4\,\kms \citep{lockman89}. The solid white circles in the lower left corners show the 2\arcmper65 beam size. 
    The 0.8\,\kms\ emission is spatially extended, whereas the 33.4\,\kms\ emission is spatially centered on G010.331$-$0.278 with an angular extent similar to that reported in the WISE Catalog (white circles).  We conclude that the velocity of G010.331$-$0.278 is 33.4\,\kms, and the 0.8\,\kms\ component is from the DIG. \label{fig:multvel}} 
\end{figure*}

In Figure~\ref{fig:multvel} we show an example of this method using the known \hii\ region G010.331$-$00.287 from the WISE catalog. We display the GDIGS \hna\ emission for the two detected velocities: 0.8\,\kms\ and 33.4\,\kms \citep{lockman89}. The emission centered at the 0.8\,\kms\ velocity is diffuse and does not match the expected morphology. The emission centered at the 33.4\,\kms\ velocity matches the expected \hii\ region morphology, and we therefore treat this as the ``correct'' \hii\ region velocity.

\begin{figure*}
    \centering
    \includegraphics[width=3.65in]{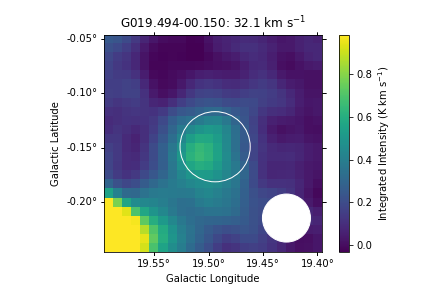}
    \hskip-0.3in
    \includegraphics[width=3.65in]{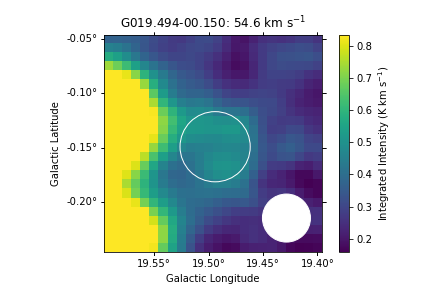}
    \caption{\hna\ Moment~0 images toward source G019.494$-$00.150. Each image is integrated over a range centered at one of the two detected RRL velocities for this source and equal in width to the corresponding FWHM. The left image is integrated over a FWHM of 16.8\,\kms with the range centered at 32.1\,\kms, and the right image is integrated over a FWHM of 15.1\,\kms with the range centered at 54.6\,\kms \citep{anderson2011}. The solid white circles in the lower right corners show the 2\arcmper65 beam size. For both velocity components we see similar emission, and therefore are unable to discern which velocity component belongs to the \hii\ region.}
    \label{fig:multvel_fail}
\end{figure*}

As a point of comparison, in Figure~\ref{fig:multvel_fail} we show an example of an unsuccessful use of the method with \hii\ region G019.494$-$00.150, also from the WISE Catalog. As in Figure~\ref{fig:multvel} we display the GDIGS \hna\ emission for the two detected velocities, this time at 32.1\,\kms\ and 54.6\,\kms \citep{anderson2011}. The emission observed from both velocity components has a good agreement with the size and location of the WISE Catalog entry; we therefore cannot determine which velocity is the ``correct'' one for the \hii\ region.

\startlongtable
\begin{deluxetable}{lrr}
\tabletypesize{\scriptsize}
\tablecaption{Multiple RRL Velocity Locations \label{tab:multvel}}
\tablewidth{0pt}
\tablehead{
\colhead{Source} &
\colhead{\hii Region $V_{\rm LSR}$} &
\colhead{Other $V_{\rm LSR}$}\\
\colhead{} &
\colhead{(\kms)} &
\colhead{(\kms)} }
\startdata
\input gdigs_multvel5.tab
\enddata
\end{deluxetable}

We compile a list of all 121 multiple-velocity \hii\ regions in the GDIGS range from the WISE Catalog.
We use GDIGS data to identify the LSR velocity of the discrete source for 35 of these 121
sources.
We give the \hii\ region velocity for these 35 sources in Table~\ref{tab:multvel}, which lists the source name, the \hii\ region RRL velocity, and the other RRL velocities detected along the line of sight.
The \hii regions G359.717$-$00.036 and G359.727$-$00.037 are in close angular proximity and have the same reported velocities. In this case, the velocities reported in the WISE Catalog were obtained by previous observations that observed both objects in the same beam. Following the methodology of the WISE Catalog, we report the same velocity components for both.
We were not able to distinguish the ``correct'' velocity for the remaining sources using our method. 
We hypothesize that the other velocities come from the DIG or from extended envelopes of other nearby \hii\ regions.

We also use GDIGS data to verify a subset of the \hii\ region velocities reported in \citet{anderson15b}. In \citet{anderson15b} the authors attempted to determine the correct velocities of 117 multiple-velocity \hii\ regions from the HRDS using new GBT RRL observations. We identified 60 \hii\ regions in that paper that fall within the GDIGS survey area and applied the method described earlier in this section. None of these 60 sources showed a disagreement with the analysis of \citet{anderson15b}.  For 45 of these \hii\ regions we successfully confirmed the previously reported velocities. We were unable to determine the velocities of the remaining 15, because we were not able to associate the morphology with that expected from the WISE Catalog.

%%%%%%%%%%%%%%%%%%%%%%%%%%%%%%%%%%
\subsubsection{Group H\,{\footnotesize II} Regions \label{sec:groupHII}}
GDIGS can provide RRL detections for ``group'' \hii\ regions, which are found in complexes with other known \hii\ regions but have not been detected previously in pointed RRL surveys. There are  $\sim\!160$ WISE Catalog group \hii\ regions in the GDIGS survey zone.  Many of these are confused with brighter nearby \hii\ regions, and so their RRL emission cannot be reliably measured. There are 40 unconfused group \hii\ regions that we examine for RRL emission. We consider group \hii\ regions unconfused if they are separated from their nearest neighbor by a beam width or more. For each of these, we extract a GDIGS \hna\ spectrum by integrating over a circle of radius given by the WISE Catalog. 
We then fit a Gaussian profile to each spectrum thus obtained. 

We give the Gaussian fit parameters for the examined group \hii\ regions in Table~\ref{tab:groups}, which lists the source name, the Galactic longitude ($\gl$) and latitude ($\gb$), the line intensity ($T_L$), the LSR velocity ($V_{LSR}$), the FWHM line width ($\Delta V$), the rms noise, and the signal-to-noise ratio (S/N). The errors in Table~\ref{tab:groups} are 1$\sigma$ uncertainties from the Gaussian fits. Of the 40 observed \hii\ regions, 18 have multiple RRL velocity components. For those sources, we apply the method described in Section~\ref{sec:multivel}. If the correct velocity can be determined, we report the correct velocity first, followed by the other velocity with an asterisk appended to the name. For sources for which we cannot determine which velocity is correct, we report both, appended with ``a'' and ``b'' in order of decreasing \hna\ intensity. All listed line parameters have a S/N of at least three, where the S/N
as defined by \citet{lenz92} is
\begin{equation}
    {\rm S/N} = 0.7 \left( \frac{T_{\rm L}}{\rm rms}\right) \left( \frac{\Delta V}{2.35} \right)^{0.5}\,,
\end{equation}
where 2.35 is the FWHM in \kms of the spectral smoothing kernel we used when generating the spectra.

We expect to detect FWHM line widths between $\sim 10$\,\kms and $\sim 35$\,\kms \citep{anderson11}.  Line widths with values lower than 10\,\kms\ imply electron temperatures $<1100$\,K \citep{anderson18}. 
Line widths over $\sim 35$\,\kms are very unlikely to be single components, so those reported here are likely the result of blended components.

\begin{deluxetable*}{lrrrcrccccr}
\tabletypesize{\scriptsize}
\tablecaption{GDIGS \hna\ RRL Parameters from WISE Catalog Group \hii\ Regions}
\tablewidth{0pt}
\tablehead{
\colhead{Source\tablenotemark{$a$}} &
\colhead{\gl} &
\colhead{\gb} & 
\colhead{$T_L$} &
\colhead{$\sigma T_L$} &
\colhead{$V_{LSR}$} &
\colhead{$\sigma V_{LSR}$} &
\colhead{$\Delta V$} &
\colhead{$\sigma \Delta V$} &
\colhead{rms} &
\colhead{S/N}
\\
\colhead{} &
\colhead{($\arcdeg$)} &
\colhead{($\arcdeg$)} &
\colhead{(mK)} &
\colhead{(mK)} &
\colhead{(\kms)} &
\colhead{(\kms)} &
\colhead{(\kms)} &
\colhead{(\kms)} &
\colhead{(mK)} &
\colhead{}
}
\startdata
\input gdigs_groups9.tab
\enddata
\label{tab:groups}
\tablenotetext{a}{Source names for HII regions with multiple detected hydrogen RRL components are appended with ``a'' and ``b'' in order of decreasing \hna\ RRL intensity. Source names for additional velocity components for sources with successfully determined ``correct'' velocities are appended with an asterisk.}
\end{deluxetable*}

%%%%%%%%%%%%%%%%%%%%%%%%%%%%%%%%%%%%%%
\subsubsection{H\,{\footnotesize II} Region Candidates}
Within the GDIGS survey range there are $\sim\!400$ WISE Catalog \hii\ region candidates that have spatially coincident radio continuum and MIR emission but lack a spectroscopic ionized gas detection.    
Since the GDIGS area has been searched several times by sensitive, pointed RRL surveys \citep[e.g.,][]{anderson11, L_Anderson_2015a}, new \hii\ region detections using GDIGS are expected only for
candidate sources that were skipped in previous surveys.
We examine these $\sim\! 400$ locations and find 27 \hii\ region candidates that are detected in the GDIGS \hna\ data and are unconfused by the RRL emission from nearby bright \hii\ regions, using the same criteria as in Section \ref{sec:groupHII}. We integrate GDIGS \hna\ data over the extent of the sources defined in the WISE catalog and fit Gaussian components. We give the RRL parameters in Table~\ref{tab:candidates}.
Eight of the nebulae are best fit with two Gaussian components. For these sources with multiple velocity components, we again apply the method described in Section~\ref{sec:multivel}. We use the same table format as that of Table~\ref{tab:groups} and the same S/N requirement as described in Section~\ref{sec:groupHII}. 
Two candidates, G037.319+00.162 and G037.320+00.168, have low S/N values and appear to have blended peaks. Blended peaks occur where two emission peaks that are broad enough and close enough in velocity appear to merge together into one broad, flat-topped peak. We fit these candidates with two Gaussian components each.

\begin{deluxetable*}{lrrrcrccccr}
\tabletypesize{\scriptsize}
\tablecaption{GDIGS \hna\ RRL Parameters from WISE Catalog Candidate \hii\ Regions}
\tablewidth{0pt}
\tablehead{
\colhead{Source\tablenotemark{a}} &
\colhead{\gl} &
\colhead{\gb} & 
\colhead{$T_L$} &
\colhead{$\sigma T_L$} &
\colhead{$V_{LSR}$} &
\colhead{$\sigma V_{LSR}$} &
\colhead{$\Delta V$} &
\colhead{$\sigma \Delta V$} &
\colhead{rms} &
\colhead{S/N}
\\
\colhead{} &
\colhead{($\arcdeg$)} &
\colhead{($\arcdeg$)} &
\colhead{(mK)} &
\colhead{(mK)} &
\colhead{(\kms)} &
\colhead{(\kms)} &
\colhead{(\kms)} &
\colhead{(\kms)} &
\colhead{(mK)} &
\colhead{}
}
\startdata
\input gdigs_candidates11.tab

\enddata
\label{tab:candidates}
\tablenotetext{a}{Source names for HII regions with multiple detected hydrogen RRL components are appended with ``a'' and ``b'' in order of decreasing \hna\ RRL intensity. Source names for additional velocity components for sources with successfully determined ``correct'' velocities are appended with an asterisk.}
\end{deluxetable*}

%%%%%%%%%%%%%%%%%%%%%%%%%%%%%%%%%%%%%%%%%%%%%%%%%%
\subsubsection{Radio-quiet H\,{\footnotesize II} Region Candidates}
Within the GDIGS survey range there are $\sim\!800$ WISE Catalog radio quiet \hii\ regions.  These nebulae have the characteristic MIR morphology of \hii\ regions, but have no detected radio continuum emission in any extant survey.  Given the sensitivity of GDIGS, we therefore do not expect many \hna\ RRL detections.  In the Galactic longitude range $17.5\degree > \ell > -5\degree$, however, the quality of public radio continuum data suitable for identifying \hii\ regions is poor so the pointed surveys may not have identified all detectable targets. In this range the GDIGS data may be able to provide RRL detections for some radio-quiet \hii\ regions.

We measure RRL emission from 21 out of the 800 WISE Catalog radio quiet \hii\ regions by integrating GDIGS \hna\ data over the extent of the sources defined in the WISE catalog.  All but one of these are in the zone $17.5\degree > \ell > -5\degree$. 
Four of these nebulae are best fit with two Gaussian components, and one with three Gaussian components.  
We give the \hna\ RRL parameters in Table~\ref{tab:rq_candidates}, with the same table format as that of Table~\ref{tab:groups} and the same S/N requirement as described in Section~\ref{sec:groupHII}.

\begin{deluxetable*}{lrrrcrccccr}
\tabletypesize{\scriptsize}
\tablecaption{GDIGS \hna\ RRL Parameters from WISE Catalog Radio-quiet \hii\ Regions}
\tablewidth{0pt}
\tablehead{
\colhead{Source\tablenotemark{a}} &
\colhead{\gl} &
\colhead{\gb} & 
\colhead{$T_L$} &
\colhead{$\sigma T_L$} &
\colhead{$V_{LSR}$} &
\colhead{$\sigma V_{LSR}$} &
\colhead{$\Delta V$} &
\colhead{$\sigma \Delta V$} &
\colhead{rms} &
\colhead{S/N}
\\
\colhead{} &
\colhead{($\arcdeg$)} &
\colhead{($\arcdeg$)} &
\colhead{(mK)} &
\colhead{(mK)} &
\colhead{(\kms)} &
\colhead{(\kms)} &
\colhead{(\kms)} &
\colhead{(\kms)} &
\colhead{(mK)} &
\colhead{}
}
\startdata
\input gdigs_candidatesrq7.tab
\enddata
\label{tab:rq_candidates}
\tablenotetext{a}{Source names for HII regions with multiple detected hydrogen RRL components are appended with ``a'' and ``b'' in order of decreasing \hna\ RRL intensity. Source names for additional velocity components for sources with successfully determined ``correct'' velocities are appended with an asterisk.}
\end{deluxetable*}

%%%%%%%%%%%%%%%%%%%%%%%%%%%%%%%%%%%%%%%%%
\subsection{New RRL Detections} 
We discover 45 discrete sources in GDIGS that are not associated with objects in the WISE Catalog.
 
Some \hii\ region candidates may have been missed by the WISE Catalog because their MIR emission is especially faint or because it was confused with other \hii\ regions in close proximity.
If the locations of discrete RRL emission coincide with MIR emission characteristic of \hii\ regions, we categorize them as newly-discovered \hii\ regions; if not, we characterize them as discrete emission zones. All WISE Catalog objects have coincident MIR emission, so the lack of coincident MIR is an important difference. 

In addition, we identify several locations of RRL emission of particular interest because of their anomalously high LSR velocities. Because of these unusual velocities, we analyze this last category separately from the rest of the \hii\ regions and discrete emission zones.

%%%%%%%%%%%%%%%%%%%%%%%%%%%%%%%%%%%%%%%%%%
\subsubsection{Newly Discovered H\,{\footnotesize II} Regions \label{sec:newHII}}
There are eight compact locations of GDIGS RRL emission that are spatially coincident with MIR emission but are not in the WISE Catalog. 
We hypothesize that these are \hii\ regions that were likely missed in the WISE Catalog because their MIR emission is faint or confused with that of other nearby regions. Of particular note is the relative faintness of these regions' 8\,\micron\ emission compared to that of other \hii regions. In all other respects these objects look like previously known \hii regions, and we do not have enough information to determine whether the deficiency in 8\,\micron\ emission reflects an inherent difference significant enough to put these objects into a different class. We measure RRL emission from these new \hii\ regions by integrating GDIGS \hna\ data over their extents defined visually from 24\,\micron\ MIR data. 
The RRL emission from three of these nebulae is best fit with two Gaussian components, and one with three Gaussian components.
We characterize the \hna\ GDIGS emission from these locations in Table~\ref{tab:new}. The table format is the same as that of Table~\ref{tab:groups}, with the addition of the region radius. We use the same method for determining the correct velocity as described in Section~\ref{sec:multivel} and the same S/N requirement as described in Section~\ref{sec:groupHII}.

\begin{deluxetable*}{lrrrrcrccccr}
\tabletypesize{\scriptsize}
\tablecaption{GDIGS \hna\ RRL Parameters from New \hii\ Regions}
\tablewidth{0pt}
\tablehead{
\colhead{Source\tablenotemark{a}} &
\colhead{\gl} &
\colhead{\gb} & 
\colhead{Radius} &
\colhead{$T_L$} &
\colhead{$\sigma T_L$} &
\colhead{$V_{LSR}$} &
\colhead{$\sigma V_{LSR}$} &
\colhead{$\Delta V$} &
\colhead{$\sigma \Delta V$} &
\colhead{rms} &
\colhead{S/N}
\\
\colhead{} &
\colhead{($\arcdeg$)} &
\colhead{($\arcdeg$)} &
\colhead{(\arcsec)} & 
\colhead{(mK)} &
\colhead{(mK)} &
\colhead{(\kms)} &
\colhead{(\kms)} &
\colhead{(\kms)} &
\colhead{(\kms)} &
\colhead{(mK)} &
\colhead{}
}
\startdata
\input gdigs_newregions7.tab
\enddata
\label{tab:new}
\tablenotetext{a}{Source names for HII regions with multiple detected hydrogen RRL components are appended with ``a'' and ``b'' in order of decreasing \hna\ RRL intensity. Source names for additional velocity components for sources with successfully determined ``correct'' velocities are appended with an asterisk.}
\end{deluxetable*}

%%%%%%%%%%%%%%%%%%%%%%%%%%%%%%%%%%%%%%%%%%%%%%%%%%
\subsubsection{Discrete Emission Zones (DEZs)} \label{compact_dig}
We discover 30 locations of GDIGS \hna\ emission in the Moment~0 maps that lack the spatially coincident MIR emission expected of \hii regions and also lack the expected radio continuum emission of supernova remnants (SNRs). We call these 30 locations discrete emission zones (DEZs). MIR emission is present for all known Galactic \hii\ regions, including those discovered in older surveys that targeted radio continuum emission peaks \citep[e.g.][]{lockman89}. We likewise expect MIR emission from planetary nebulae \citep{anderson12a}.
Most SNRs lack MIR emission \citep{Fuerst87}, and because they are non-thermal radio sources, are not expected to be strong in RRL emission \citep{liu19}. Any RRL emission from SNRs is expected to occur at frequencies lower than those observed in GDIGS.

With available data, we cannot conclusively 
categorize the emission sources,
and here we only characterize their \hna\ emission properties.
We measure RRL emission from these locations by integrating GDIGS \hna\ data over their extents defined by visually fitting circles using the GDIGS Moment~0 map. 
Of the 30 nebulae, 
11 are best fit with two Gaussian components, and one with three Gaussian components. 
We give the GDIGS \hna\ line parameters locations in Table~\ref{tab:new_dig}. The table format is the same as that of Table~\ref{tab:new}. We use the same method for determining the correct velocity as described in Section~\ref{sec:multivel} and the same S/N requirement as described in Section~\ref{sec:groupHII}.
For these DEZs, the ``correct'' velocity is the one that best matches the emission seen in the initial Moment~0 maps integrated over all velocities.

\begin{deluxetable*}{lrrrrcrccccr}
\tabletypesize{\scriptsize}
\tablecaption{GDIGS \hna\ RRL Parameters from Discrete Emission Zones}
\tablewidth{0pt}
\tablehead{
\colhead{Source\tablenotemark{a}} &
\colhead{\gl} &
\colhead{\gb} &
\colhead{Radius} &
\colhead{$T_L$} &
\colhead{$\sigma T_L$} &
\colhead{$V_{LSR}$} &
\colhead{$\sigma V_{LSR}$} &
\colhead{$\Delta V$} &
\colhead{$\sigma \Delta V$} &
\colhead{rms} &
\colhead{S/N}
\\
\colhead{} &
\colhead{($\arcdeg$)} &
\colhead{($\arcdeg$)} &
\colhead{(\arcsec)} &
\colhead{(mK)} &
\colhead{(mK)} &
\colhead{(\kms)} &
\colhead{(\kms)} &
\colhead{(\kms)} &
\colhead{(\kms)} &
\colhead{(mK)} &
\colhead{}
}
\startdata
\input gdigs_newregionsdig15.tab
\enddata
\label{tab:new_dig}
\tablenotetext{a}{Source names for HII regions with multiple detected hydrogen RRL components are appended with ``a'' and ``b'' in order of decreasing \hna\ RRL intensity. Source names for additional velocity components for sources with successfully determined ``correct'' velocities are appended with an asterisk.}
\end{deluxetable*}

%%%%%%%%%%%%%%%%%%%%%%%%%%%%%%%%%%%%%%%%%%%
\subsubsection{Anomalous Velocity Features}
\label{sec:AVFs}
We identify 10 emission zones in the GDIGS \hna\ Moment~0 maps that are not known to be \hii\ regions  
and have anomalously high velocities compared to that of other \hii\ regions at the same Galactic longitudes. We call such emission zones anomalous velocity features (AVFs) and estimate their positions and radii by visually fitting circles to the emission in the GDIGS \hna\ Moment~0 map.

For each of the 10 AVFs, we extract a GDIGS \hna\ spectrum by integrating over the extent of the source. 
To each such spectrum we then fit a Gaussian profile. 

We designate the 10 identified AVFs by their Galactic latitude and longitude. 
We provide the Gaussian fit parameters in Table~\ref{tab:anomalies}. The table format is the same as that of Table~\ref{tab:new}. We use the same method for determining the correct velocity as described in Section~\ref{sec:multivel} and the same S/N requirement as described in Section~\ref{sec:groupHII}.

\begin{deluxetable*}{lrrcccccccccr}
\tabletypesize{\scriptsize}
\tablecaption{Anomalous Velocity Features (AVFs)
\label{tab:anomalies}} 
\tablewidth{0pt}
\tablehead{
\colhead{Source} &
\colhead{\gl} &
\colhead{\gb} & 
\colhead{Radius} &
\colhead{$T_{L}$} &
\colhead{$\sigma T_{L}$} &
\colhead{$V_{LSR}$} &
\colhead{$\sigma V_{LSR}$} &
\colhead{$\Delta V$} &
\colhead{$\sigma \Delta V$} & 
\colhead{rms} & 
\colhead{S/N}
\\
\colhead{} &
\colhead{($\arcdeg$)} &
\colhead{($\arcdeg$)} &
\colhead{(\arcsec)} &
\colhead{(mK)} &
\colhead{(mK)} &
\colhead{(\kms)} & 
\colhead{(\kms)} &
\colhead{(\kms)} &
\colhead{(\kms)} & 
\colhead{(\kms)} &
\colhead{}
}
\startdata
\input gdigs_avf2.tab
\enddata

\end{deluxetable*}

%%%%%%
\section{Discussion}

\subsection{WISE Catalog \hii\ regions}
Because \hii regions are formed by short-lived massive stars, a sufficiently large sample of \hii regions with known distances can be used to measure the current properties of Galactic massive star formation \citep{anderson18}. The present work adds \hii\ regions available to such studies by confirming 27 \hii region candidates, 21 radio-quiet \hii\ regions, and 40 group \hii\ regions as known \hii\ regions, adding eight previously unreported \hii\ regions, 
and confirming the velocities of 35 previously known \hii\ regions. These numbers are tabulated in Table~\ref{tab:vlsr_nums}.
Together, these 131 new velocity measurements represent a  5.6\% increase in the number of WISE Catalog objects with a known velocity.

\begin{deluxetable}{lr}
\tabletypesize{\scriptsize}
\tablecaption{New RRL Detections by Category}
\label{tab:vlsr_nums}
\tablewidth{0pt}
\tablehead{
\colhead{Source Category} &
\colhead{Number} 
}
\startdata
WISE Catalog Objects&\\
~~~~~~\hii\ Region Candidates & 27 \\
~~~~~~Radio-Quiet \hii\ Region Candidates & 21 \\
~~~~~~Group \hii\ Regions & 40 \\
~~~~~~Known \hii\ Regions & 35 \\
New \hii\ Regions & 8 \\
DEZs & 30 \\
AVFs & 10
\enddata
\end{deluxetable}

\begin{figure}
    \centering
    \includegraphics[width=3.2in]{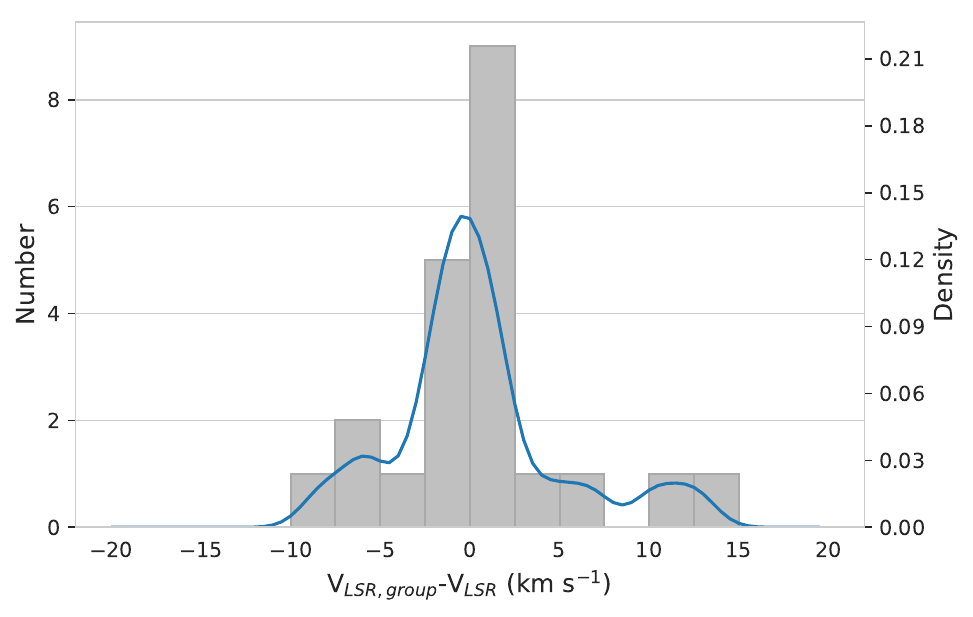}
    \caption{Histogram of differences between the measured GDIGS RRL velocity of 22 group \hii\ regions and the group velocity reported in the WISE Catalog, overlaid with the kernel density estimate (KDE) curve for the same. The 18 \hii\ regions with multiple velocity components have been excluded. The group \hii\ regions all have velocities within 20\,\kms\ of the velocities of their \hii\ region complexes.}
    \label{fig:group}
\end{figure}

The GDIGS RRL velocities of group \hii\ regions indicate that they are for the most part correctly associated with known \hii\ regions. Of the 40 observed group \hii\ regions, 38 have one observed LSR velocity within 10\,\kms of the ``group velocity'' (the velocity measured for the \hii\ region complex) reported in the WISE Catalog; 
all have one component within 20\,\kms. 
In Figure~\ref{fig:group}, we plot the differences between the measured GDIGS group \hii\ region RRL velocity and the group velocity as reported in the WISE Catalog.
\hii\ regions with multiple observed velocities are excluded from this plot.
There is a strong correlation between the two velocities, indicating that the original group associations for this sample were largely correct.  

\subsection{DEZs}
Ionized gas comprises a significant portion of the Galactic ISM, accounting for more than 20\% of the Milky Way's gas mass \citep{reynolds91a}. Despite this, current understanding of the ionized component of the ISM is far from complete. The identification and characterization here of 30 discrete zones of RRL emission suggests that the Galactic DIG may consist in part of compact emission zones with sizes of a few arcminutes.

\begin{figure}
    \centering
    \includegraphics[width=3.65in]{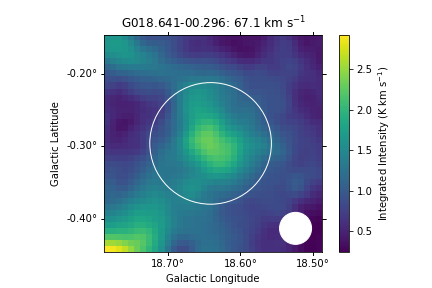}
    \caption{Example \hna\ Moment~0 map of the DEZ G018.641$-$0.296. The image is integrated over a range centered at the velocity detected for this DEZ, 67.1\,\kms\, and equal in width to the detected FWHM, 36.8\,\kms. The white circle marks the estimated extent of the DEZ. The filled white circle in the lower right corner shows the 2\arcmper65 GDIGS beam size.}
    \label{fig:DEZ}
\end{figure}

The \hna\ RRL morphology of the DEZs is generally centrally-peaked. Some DEZs have nearly circular RRL emission, some are oblong, and some are irregular in shape. In Figure~\ref{fig:DEZ} we show an example Moment~0 map of a typical DEZ, G018.641$-$00.296.

We examine 21\,\cm radio continuum emission from the Very Large Array (VLA) Galactic Plane Survey (VGPS) at the locations of the 11 DEZs that fall within the VGPS survey area and found identifiable radio continuum emission for G026.584+00.163 and G027.128+00.010. 

One of the DEZs, G012.508$-$00.127, has very faint 24\,\micron\ emission near its center, but lacks the 8\,\micron\ emission we also require from \hii\ regions.

\subsection{AVFs}
The ten objects that we call AVFs merit further study. They do not fit neatly into current models of gas behavior in the Galactic disk \citep{reid19}, and their existence may require an update to those models.  
We plan additional observations 
in order to better understand the origins of these objects and the reason for their high velocities. 

To investigate the nature of these features, we compare GDIGS RRL data with the MIR data described in Section \ref{sec:MIR_data}, as well as with \hi\ from the Galactic All-Sky Survey \citep[GASS;][]{mcclure-griffiths09}, and with \cor\ $J=2\rightarrow 1$ spectra from the Structure, Excitation, and Dynamics of the Inner Galactic ISM survey \citep[SEDIGISM;][]{schuller20} and the $J=1\rightarrow 0$ from the Galactic Ring Survey \citep[GRS;][]{jackson06}. 
GASS has a spectral resolution of 0.82\,\kms\ and a sensitivity of 57\,\mK.  
SEDIGISM has 1.0\,\K\ 1$\sigma$ sensitivity, 30\,\arcsec\ angular resolution, and 0.35\,\kms\ spectral resolution.  The GRS has 46\,\arcsec\ angular resolution, 0.2\,\kms\ spectral resolution, and 0.4\,\K\ sensitivity. 
We compare our RRL observations to \cor\ data from these surveys to search for any associated molecular gas, which is commonly found for Galactic \hii\ regions.
\citep{anderson09a}.

\begin{figure*}
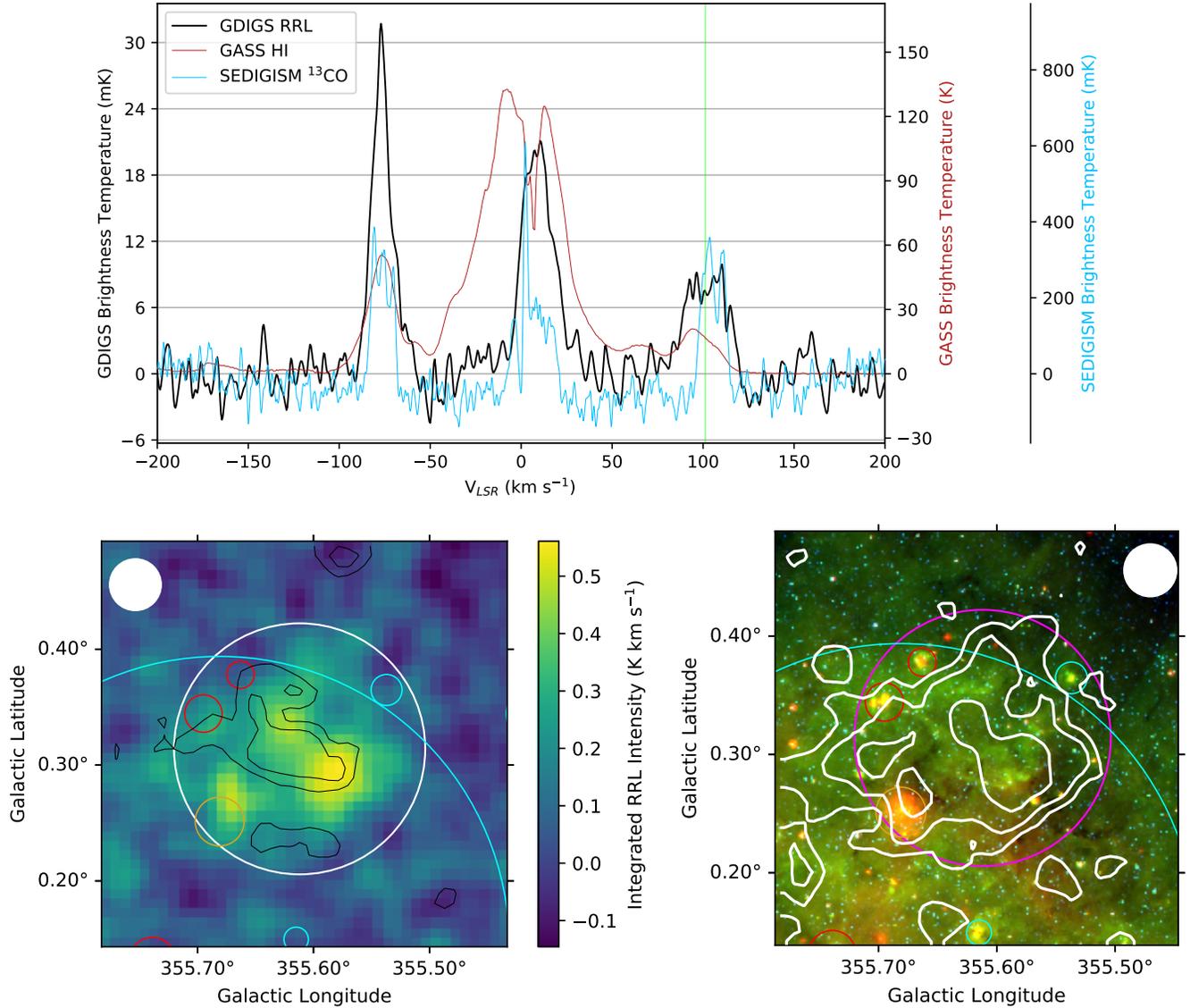

    \centering
    \includegraphics[width=6in]{hv1_l355.61_b0.314_v101.23_r390_smoothed2.35kms_combined_spectra.pdf}
    \includegraphics[width=3.89in]{hv1_l355.612_b0.314_v101.23_r390_sm2_b2_f1.75_combined_mom0.pdf}
   \includegraphics[width=3.11in]{hv1_l355.612_b0.314_v101.23_r390_percentile98_individualnorm_NIR_3color.pdf}
   \caption{Top: Spectra for AVF G355.612+00.314. GDIGS RRL data are shown in black, GASS \hi\ data in red, and SEDIGISM \cor\ data in blue. The GDIGS data are spectrally Gaussian-smoothed with a FWHM of 2.35\,\kms. The vertical green line marks the estimated velocity of the AVF. 
   Bottom left: Moment~0 map for the same AVF. The color map displays the GDIGS RRL Moment~0 integrated intensity, while the black contours display the SEDIGISM \cor Moment~0 integrated intensity. As in Section \ref{sec:multivel}, the Moment~0 integration is carried out over a velocity range equal to the FWHM and centered on the LSR velocity. The \cor\ data are spatially Gaussian-smoothed with a FWHM of 19\,\arcsec. The contour levels are at 2$\sigma$ and 3.5$\sigma$, where $\sigma$ is the standard deviation of the \cor Moment~0 intensity across the circle. The white circle marks the estimated extent of the AVF. The filled white circle in the upper left shows the 2\arcmper65 GDIGS beam size. Red circles mark known \hii\ regions,  
   orange circles mark radio-quiet \hii\ regions, and cyan circles mark \hii\ region candidates, all from the WISE Catalog. Of the two known \hii\ regions in the area, the northern one has a velocity of $-$2.6\,\kms\, while the other has velocity components of 3.0\,\kms\ and $-$79.1\,\kms. We do not think either is associated with the AVF.
   Bottom right: Example three-color composite image of the same AVF made from Spitzer MIR data; 24\,\micron\ emission is represented in red, 8\,\micron\ emission is represented in green, and 3.6\,\micron\ emission is represented in blue. White contours display the GDIGS RRL Moment~0 intensity from the left panel. The contours on this map are at $\sigma$, 1.8$\sigma$, and 3.1$\sigma$, where $\sigma$ is the standard deviation of the RRL Moment~0 intensity across the circle. A magenta circle marks the estimated extent of the AVF, and red, orange, and cyan circles mark \hii\ regions, radio-quiet \hii\ regions, and \hii\ region candidates from the WISE catalog. The filled white circle in the upper right again shows the 2\arcmper65 GDIGS beam size. A figure set containing maps and spectra for the 10 AVFs discussed in Section~\ref{sec:AVFs} is available in the online version of this article.}
    \label{fig:AVF_samples}
\end{figure*}

We show sample spectra, Moment~0 maps, and MIR maps in Figure~\ref{fig:AVF_samples}. The complete figure set (30 images) can be found in the online version of this article.  Morphologically, the RRL Moment~0 maps of AVFs have areas of bright emission with adjacent diffuse emission. In half of the AVFs the bright emission is concentrated in one clump, while in the other half it is split between two to three smaller clumps connected by diffuse emission. Two of the latter group have a structure resembling a shell. In eight out of ten cases each bright area has diffuse emission in all directions that gradually decreases in intensity. In the remaining two AVFs the intensity of GDIGS emission drops off sharply in at least one direction. The individual clumps vary in shape between compact, oblong, and irregular structures. The wide variety of morphologies seen in the AVFs further confounds attempts to determine their nature and origin, and raises the possibility that the AV
Fs might not all be the same type of object. 

We compare AVF observations with MIR data to determine if the objects may be \hii\ regions, and find that none have the associated 24\,\micron emission that is characteristic of \hii\ regions. Nine have 24\,\micron\ MIR emission near to the RRL emission, but in those cases it is inconsistent with the RRL emission in spatial location or size and is often associated with other known objects. We therefore do not consider that MIR emission is associated with these nine AVFs, and correspondingly, find no evidence that AVFs are \hii\ regions. 

All ten AVF spectra show \hi\ spectrally coincident with the AVF's RRL emission, which signals the presence of associated neutral gas.
Since \hi\ emits at all allowed velocities in the Galaxy, this result is expected.
Four AVFs have \cor\ spatially coincident with the RRL emission, and
six AVF spectra show spectrally coincident \cor\ emission. For these six, however, the RRL and \cor\ emission have different spatial distributions. 
While the circle defining the limits of each AVF is drawn to fit as tightly as possible, the AVFs are not circular so the RRL emission does not completely fill the circle. In these six cases, \cor\ emission is observed inside the circle over which the spectrum is integrated, but is not spatially coincident with the RRL emission.

Eight AVFs are at least partially co-spatial with objects from the WISE Catalog.
Among these eight AVFs, there are 13 known \hii\ regions, 11 \hii\ region candidates, and 12 radio-quiet \hii\ regions fully or partially co-spatial with an AVF. Based on their relative locations,  angular sizes, and velocities we do not consider any of the catalog objects to be associated with the AVFs. 
In the case of the known \hii\ regions, we check the AVF's velocity against the velocity of the \hii\ region reported in the WISE Catalog. If they are not within 20\,\kms, we consider the AVF and \hii\ region unlikely to be associated. We also compare the position and morphology of the AVF to the reported size and position of the \hii\ region in the WISE Catalog. In all cases, either the velocity does not match, the position and morphology do not match, or both. We require all these criteria to match in order to consider the AVF and \hii\ region to be associated.
\hii\ region candidates and radio-quiet sources do not have measured velocities, and as such, we cannot make a velocity comparison between these regions and the AVFs.
The position and morphology of the AVFs do not correspond to any such sources in the WISE Catalog.

\begin{figure*}
    \centering
    \includegraphics[width=6in]{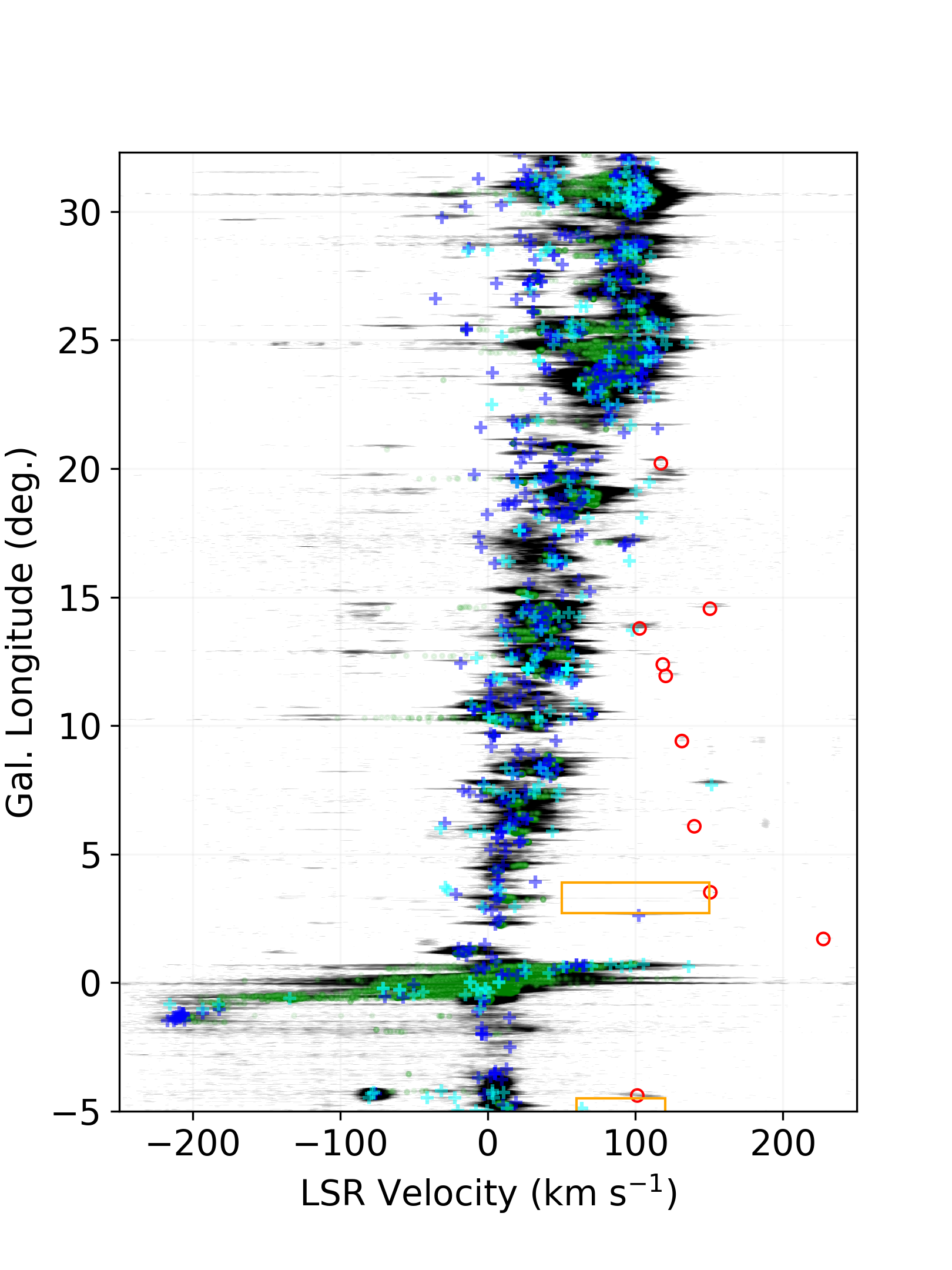}   
    \caption{Longitude-velocity (LV) diagram of the GDIGS survey data (grayscale). Known \hii\ regions from the WISE catalog are marked with crosses: blue for regions with one velocity component, cyan for each component of those with multiple velocity components. Small green circles mark the centroids of the Gaussian decompositions of the GDIGS data \citep[see][]{anderson21}. Larger red circles mark the AVFs. The orange boxes show the approximate position and dimensions of Bania's Clump~1 and~2 \citep{Bania77}.
    \label{fig:LV}}
\end{figure*}

In Figure~\ref{fig:LV} we show the location of these AVFs on a longitude-velocity (LV) diagram along with GDIGS data, and known \hii\ regions from the WISE Catalog.
We also show rectangles with the approximate longitude and velocity ranges of Bania's Clumps 1 and 2 \citep{Bania77}. These clumps are large, isolated regions of \co\ emission observed in the direction of the inner Galaxy. Both are $\sim 1$\,\degree\ wide in Galactic longitude, with Clump~1 centered at $\sim 355$\degree\ and Clump~2 centered at $\sim 3$\degree. Clump~1 spans a velocity range of $\sim 60-120$\kms\, while Clump~2 covers a range of $\sim 50-150$\kms. The velocity range of Clump~1 exceeds that allowed by circular rotation at its Galactic longitude, and therefore much of the its velocity is expected to be noncircular \citep{Bania77}. Clump~2 is notable for having the second widest velocity range of all Galactic \co\ features, with only the nuclear disk exceeding it. Bania's Clumps are observed over longitude and velocity ranges comparable to some of the AVFs, and are included for comparison.

\figsetstart
\figsetnum{5}
\figsettitle{Anomalous Velocity Features (AVFs)}

\figsetgrpstart
\figsetgrpnum{5.1}
\figsetgrptitle{G355.612+00.314}
\figsetplot{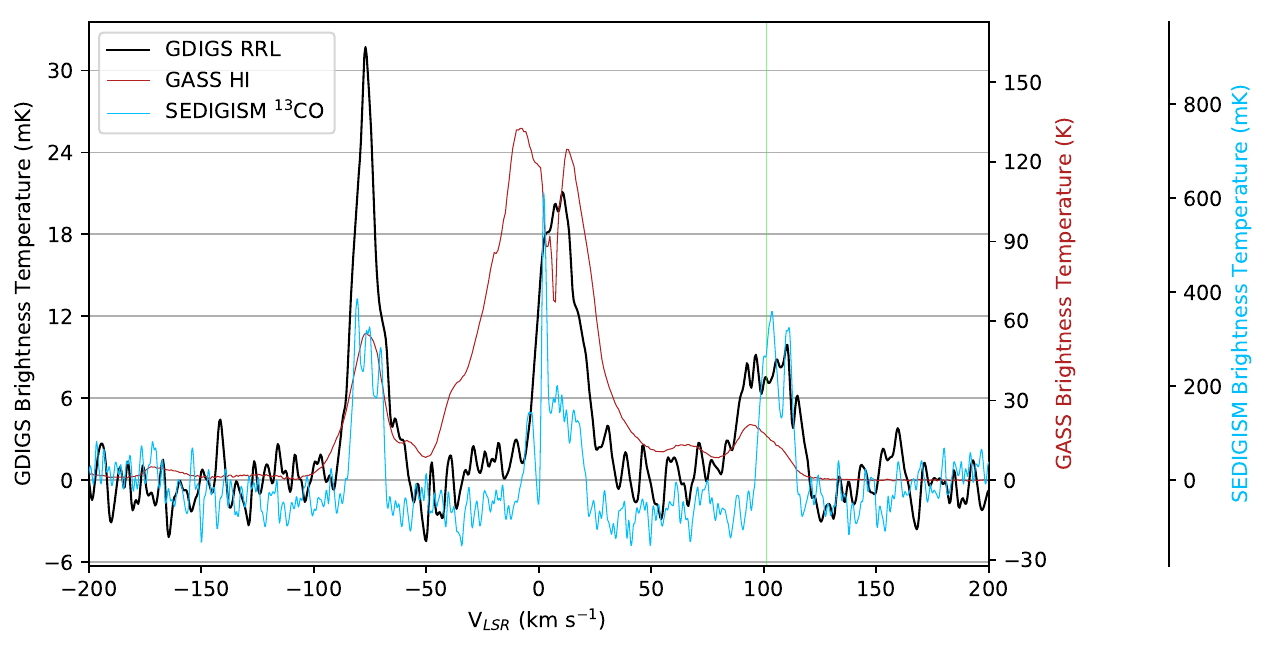}
\figsetplot{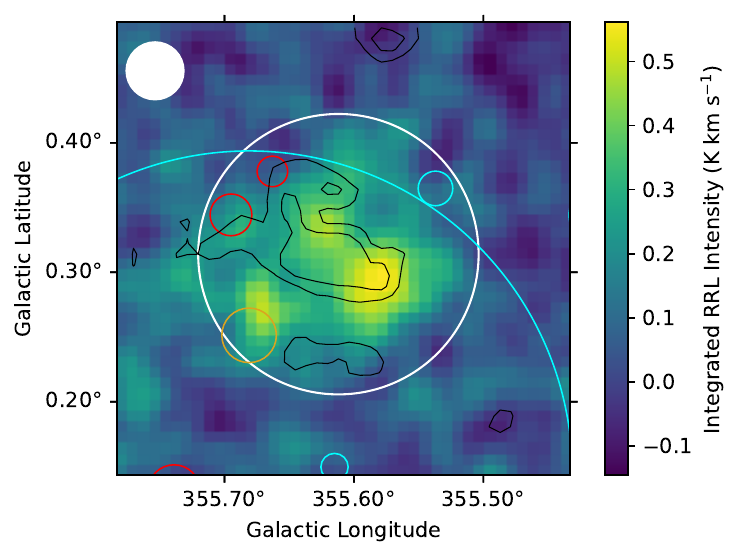}
\figsetplot{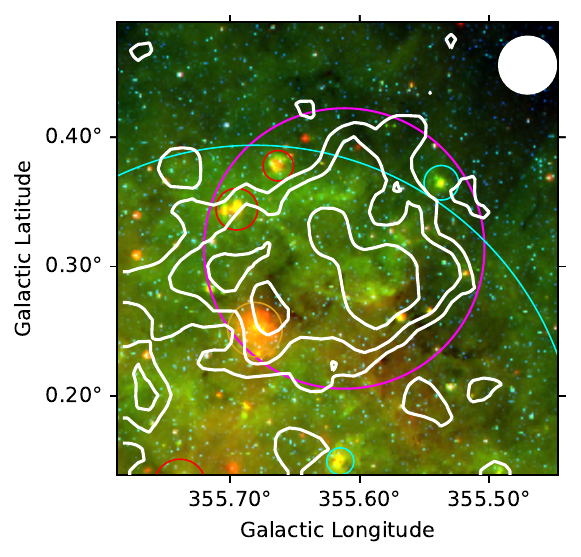}
\figsetgrpnote{The spectrum (1), RRL Moment~0 map with \cor\ contours (2), and IR image with RRL contours (3) of the AVF at Galactic coordinates 355.612+00.314. The AVF has an approximate angular radius of 390\,\arcsec and an approximate LSR velocity of 101.2\,\kms\. The GDIGS spectrum is Gaussian-smoothed with a FWHM of 5\,\kms\. \cor\ contours on the RRL map are at 2$\sigma$ and 3.5$\sigma$ and are made with data spatially Gaussian-smoothed with a FWHM of 19\,\arcsec. RRL contours on the IR image are at $\sigma$, 1.8$/sigma$, and 3.1$\sigma$. In the IR image, 24\,\micron emission is represented in red, 8\,\micron emission is represented in green, and 3.6\,\micron emission is represented in blue. The solid white circles in the corners of the RRL Moment~0 map and the IR image indicate the size of the GDIGS beam. These three images are featured as samples in the main text.}
\figsetgrpend

\figsetgrpstart
\figsetgrpnum{5.2}
\figsetgrptitle{G001.702-00.156}
\figsetplot{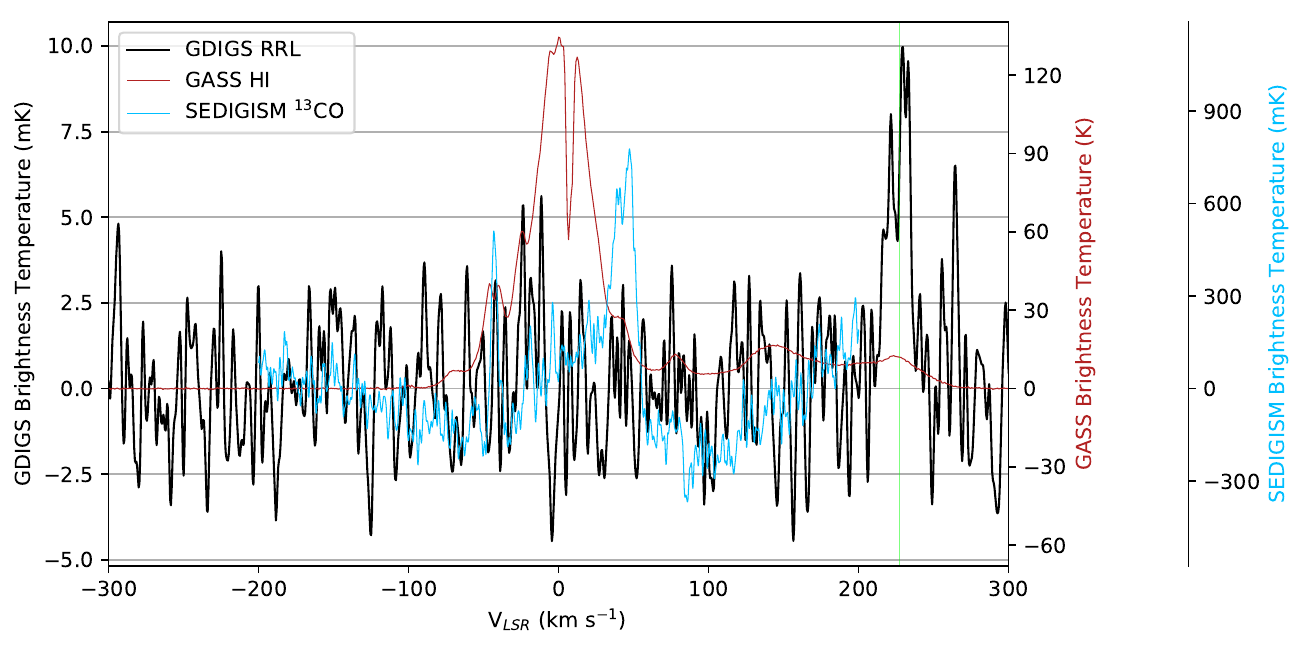}
\figsetplot{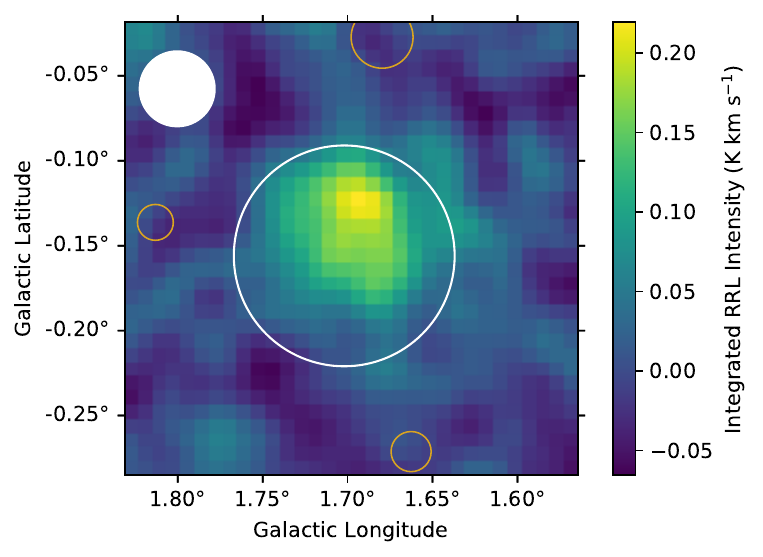}
\figsetplot{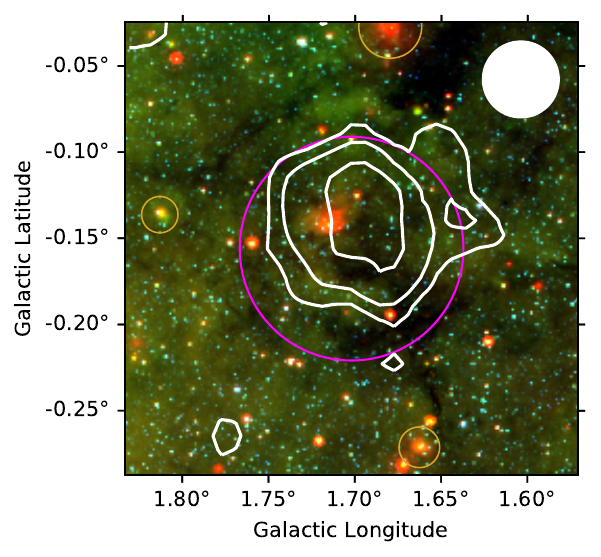}
\figsetgrpnote{The spectrum (1), RRL Moment~0 map (2), and IR image with RRL contours (3) of the AVF at Galactic coordinates 001.702-00.156. The AVF has an approximate angular radius of 234\,\arcsec and an approximate LSR velocity of 227.3\,\kms\. The GDIGS spectrum is Gaussian-smoothed with a FWHM of 5 \kms\. This AVF is beyond the velocity range of the \cor\ data, so there is none to display. RRL contours on the IR image are at $\sigma$, 1.8$\sigma$, and 3.1$\sigma$. In the IR image, 24\,\micron emission is represented in red, 8\,\micron emission is represented in green, and 3.6\,\micron emission is represented in blue. The solid white circles in the corners of the RRL Moment~0 map and the IR image indicate the size of the GDIGS beam.}
\figsetgrpend

\figsetgrpstart
\figsetgrpnum{5.3}
\figsetgrptitle{G003.518-00.040}
\figsetplot{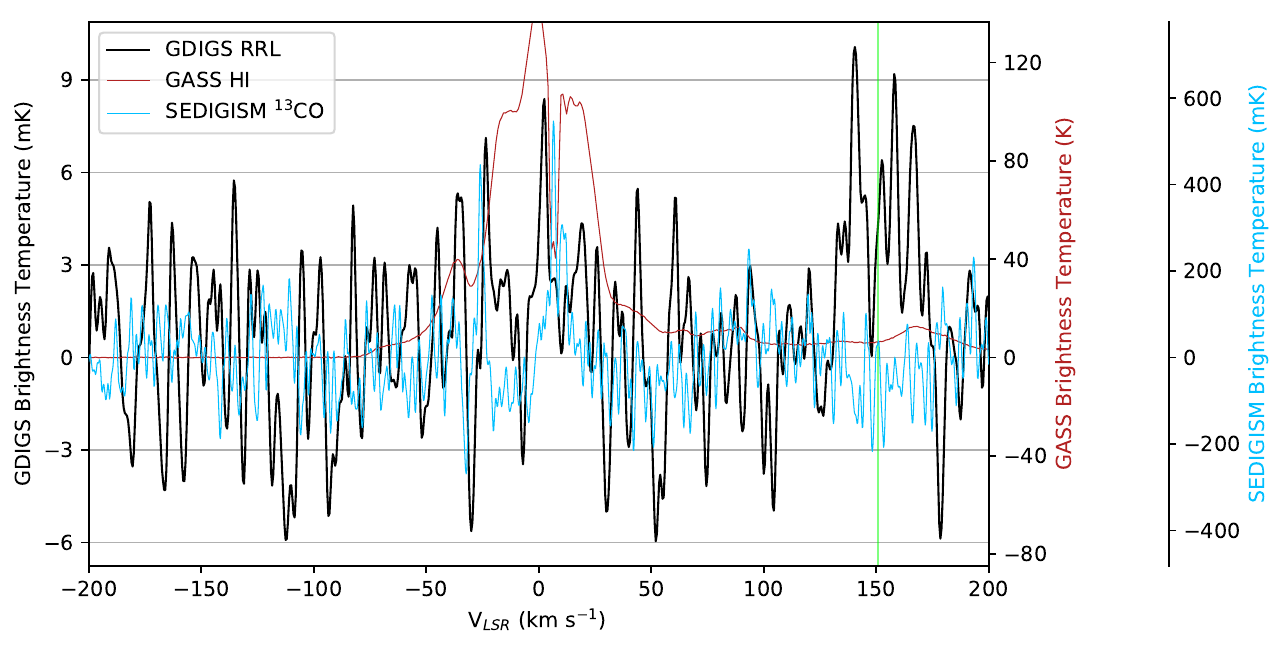}
\figsetplot{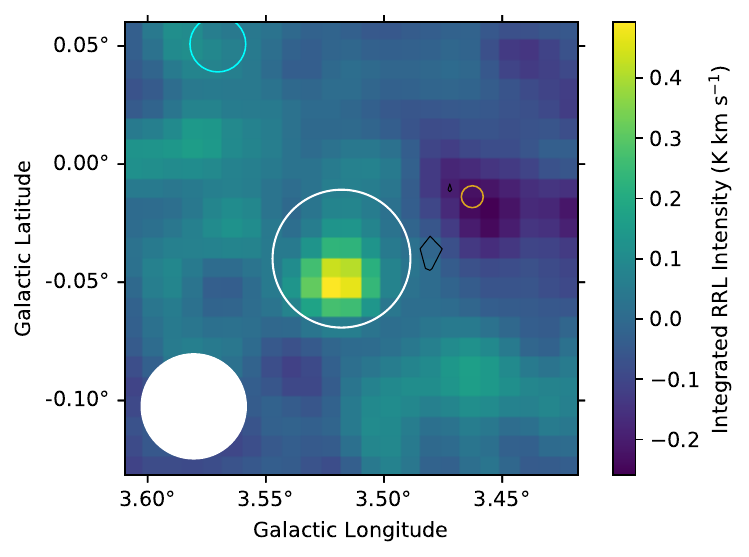}
\figsetplot{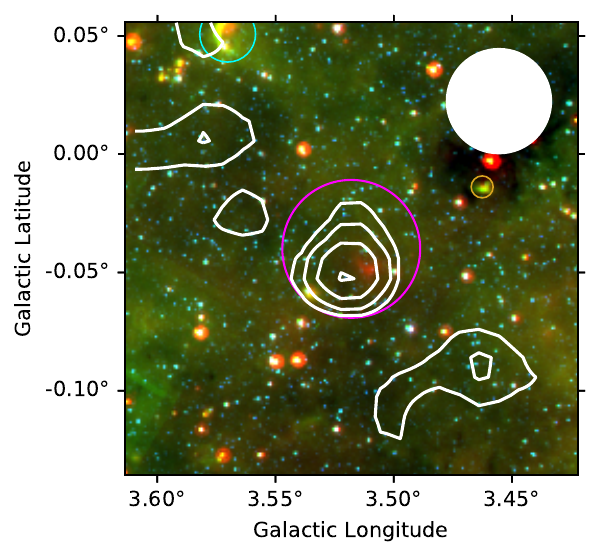}
\figsetgrpnote{The spectrum (1), RRL Moment~0 map with \cor\ contours (2), and IR image with RRL contours (3) of the AVF at Galactic coordinates 003.518-00.040. The AVF has an approximate angular radius of 143\,\arcsec and an approximate LSR velocity of 150.7\,\kms\. The GDIGS spectrum is Gaussian-smoothed with a FWHM of 5 \kms\. \cor\ contours on the RRL map are at 2$\sigma$ and are made with data spatially Gaussian-smoothed with a FWHM of 19\,\arcsec. RRL contours on the IR image are at $\sigma$, 1.8$\sigma$, 3.1$\sigma$, and 5.4$\sigma$. In the IR image, 24\,\micron emission is represented in red, 8\,\micron emission is represented in green, and 3.6\,\micron emission is represented in blue. The solid white circles in the corners of the RRL Moment~0 map and the IR image indicate the size of the GDIGS beam.}
\figsetgrpend

\figsetgrpstart
\figsetgrpnum{5.4}
\figsetgrptitle{G006.085-00.016}
\figsetplot{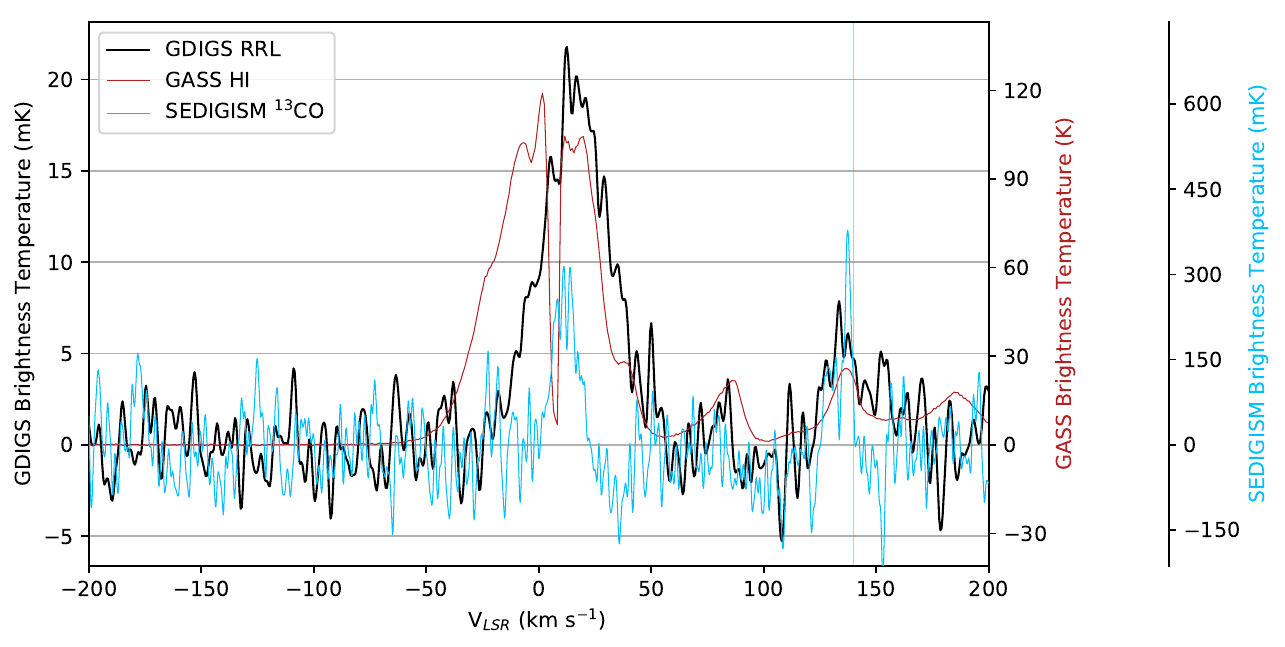}
\figsetplot{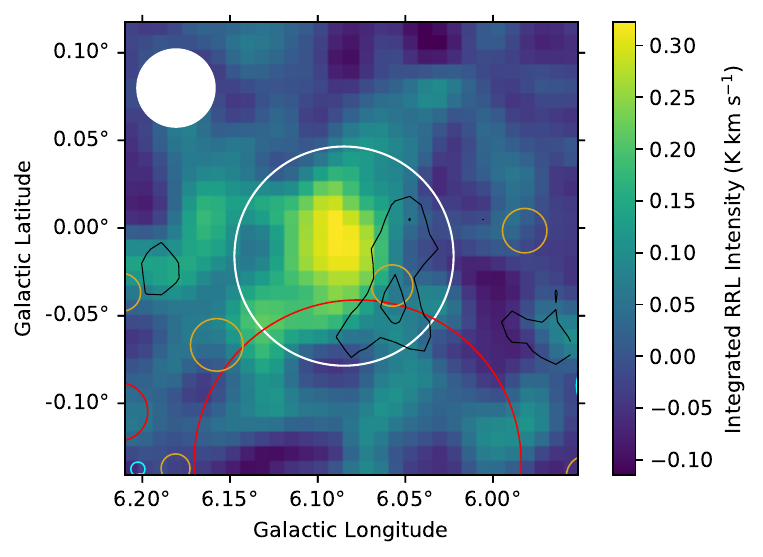}
\figsetplot{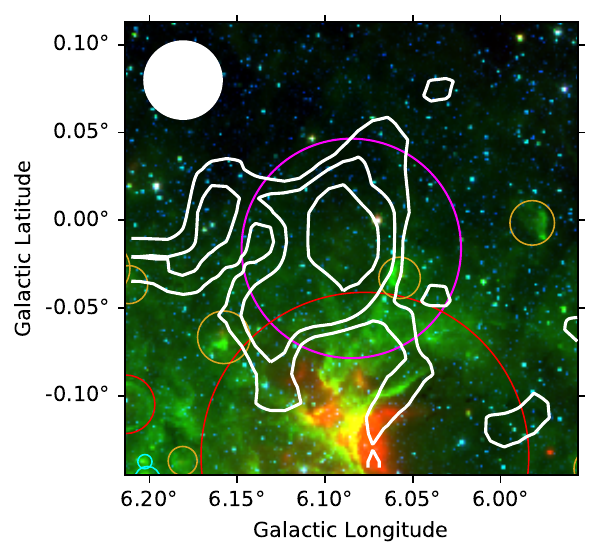}
\figsetgrpnote{The spectrum (1), RRL Moment~0 map with \cor\ contours (2), and IR image with RRL contours (3) of the AVF at galactic coordinates 006.085-00.016. The AVF has an approximate angular radius of 225\,\arcsec and an approximate LSR velocity of 139.9\,\kms\. The GDIGS spectrum is Gaussian-smoothed with a FWHM of 5 \kms\. \cor\ contours on the RRL map are at 2$\sigma$ and 3.5$\sigma$ and are made with data spatially Gaussian-smoothed with a FWHM of 19\,\arcsec. RRL contours on the IR image are at $\sigma$, 1.8$\sigma$, and 3.1$\sigma$. In the IR image, 24\,\micron emission is represented in red, 8\,\micron emission is represented in green, and 3.6\,\micron emission is represented in blue. The solid white circles in the corners of the RRL Moment~0 map and the IR image indicate the size of the GDIGS beam.}
\figsetgrpend

\figsetgrpstart
\figsetgrpnum{5.5}
\figsetgrptitle{G009.402+00.180}
\figsetplot{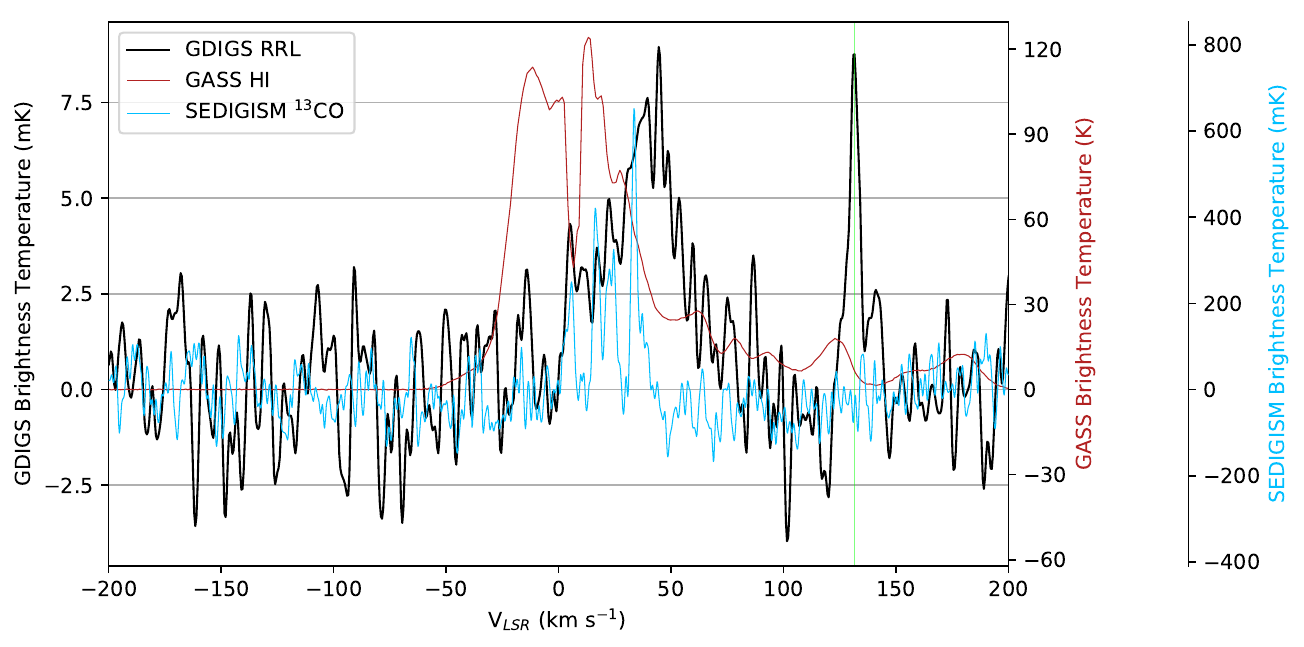}
\figsetplot{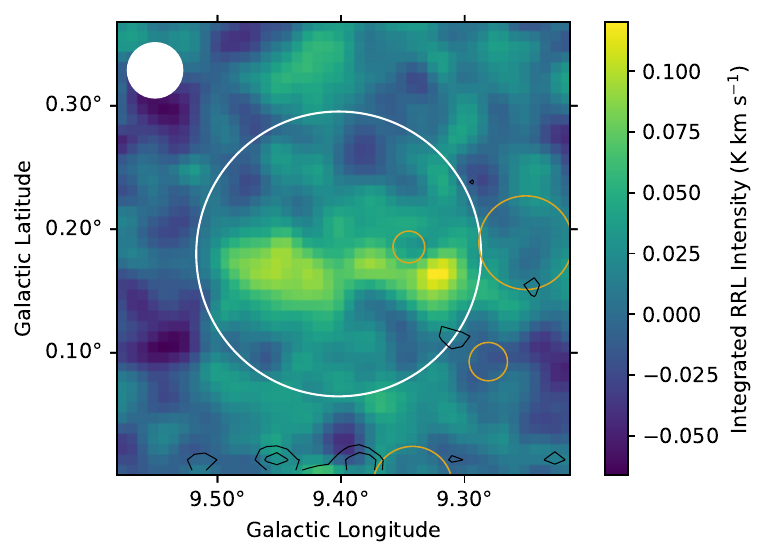}
\figsetplot{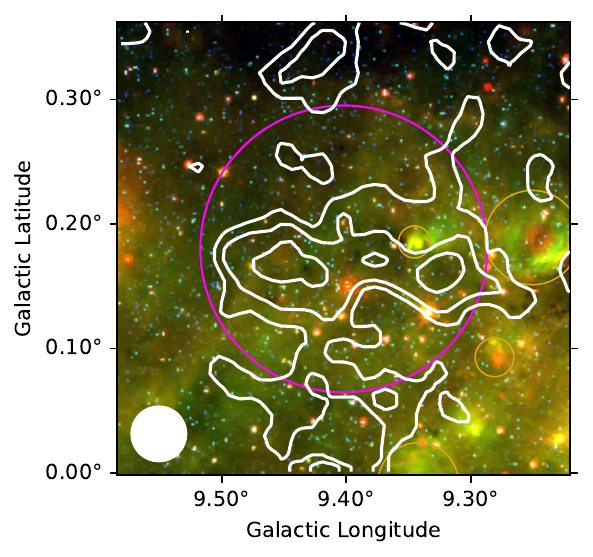}
\figsetgrpnote{The spectrum (1), RRL Moment~0 map with \cor\ contours (2), and IR image with RRL contours (3) of the AVF at galactic coordinates 009.402+00.180. The AVF has an approximate angular radius of 415\,\arcsec and an approximate LSR velocity of 131.5\,\kms\. The GDIGS spectrum is Gaussian-smoothed with a FWHM of 5 \kms\. \cor\ contours on the RRL map are at 2$\sigma$ and 3.5$\sigma$ and are made with data spatially Gaussian-smoothed with a FWHM of 19\,\arcsec. RRL contours on the IR image are at $\sigma$, 1.8$\sigma$, and 3.1$\sigma$. In the IR image, 24\,\micron emission is represented in red, 8\,\micron emission is represented in green, and 3.6\,\micron emission is represented in blue. The solid white circles in the corners of the RRL Moment~0 map and the IR image indicate the size of the GDIGS beam.}
\figsetgrpend

\figsetgrpstart
\figsetgrpnum{5.6}
\figsetgrptitle{G011.936+00.020}
\figsetplot{hv12_l11.94_b0.02_v120.52_r282_smoothed2.35kms_combined_spectra.pdf}
\figsetplot{hv12_l11.936_b0.02_v120.52_r282_sm2_b2_f1.75_combined_mom0.pdf}
\figsetplot{hv12_l11.936_b0.02_v120.52_r282_percentile98_individualnorm_NIR_3color.pdf}
\figsetgrpnote{The spectrum (1), RRL Moment~0 map with \cor\ contours (2), and IR image with RRL contours (3) of the AVF at galactic coordinates 011.936+00.020. The AVF has an approximate angular radius of 282\,\arcsec and an approximate  LSR velocity of 120.5\,\kms\. The GDIGS spectrum is Gaussian-smoothed with a FWHM of 5 \kms\. \cor\ contours on the RRL map are at 2$\sigma$, 3.5$\sigma$, and 6.1$/sigma$ and are made with data spatially Gaussian-smoothed with a FWHM of 19\,\arcsec. RRL contours on the IR image are at $\sigma$, 1.8$\sigma$, and 3.1$\sigma$. In the IR image, 24\,\micron emission is represented in red, 8\,\micron emission is represented in green, and 3.6\,\micron emission is represented in blue. The solid white circles in the corners of the RRL Moment~0 map and the IR image indicate the size of the GDIGS beam.}
\figsetgrpend

\figsetgrpstart
\figsetgrpnum{5.7}
\figsetgrptitle{G012.380-00.081}
\figsetplot{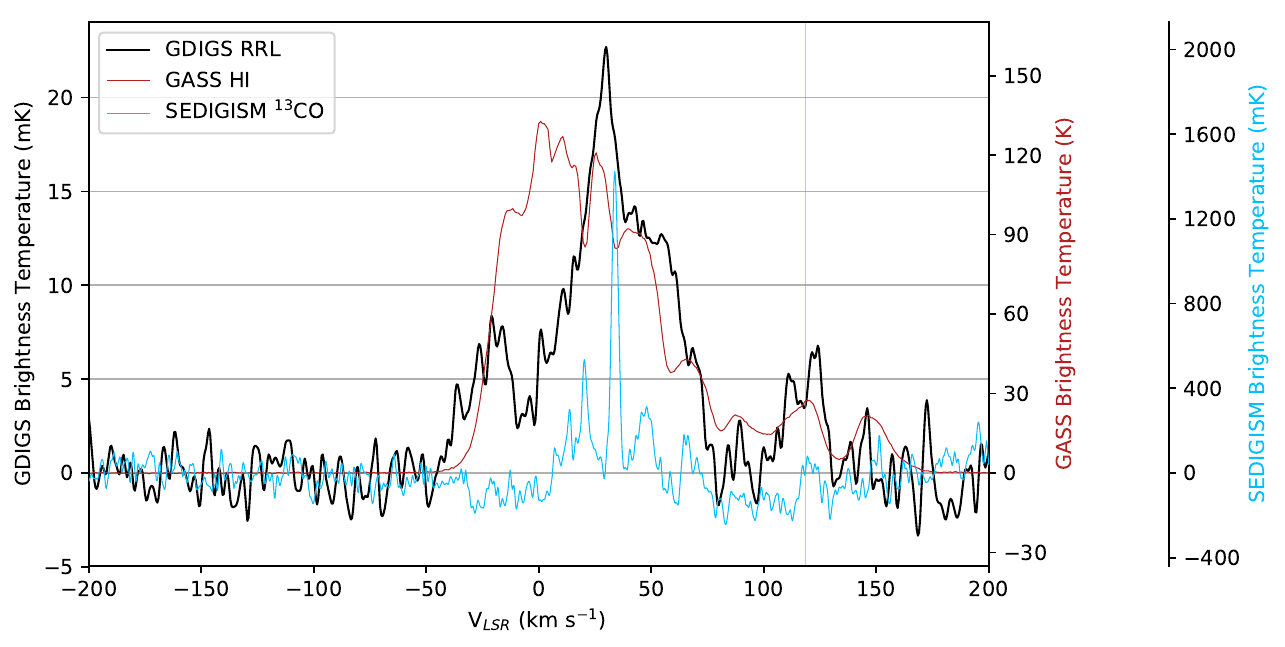}
\figsetplot{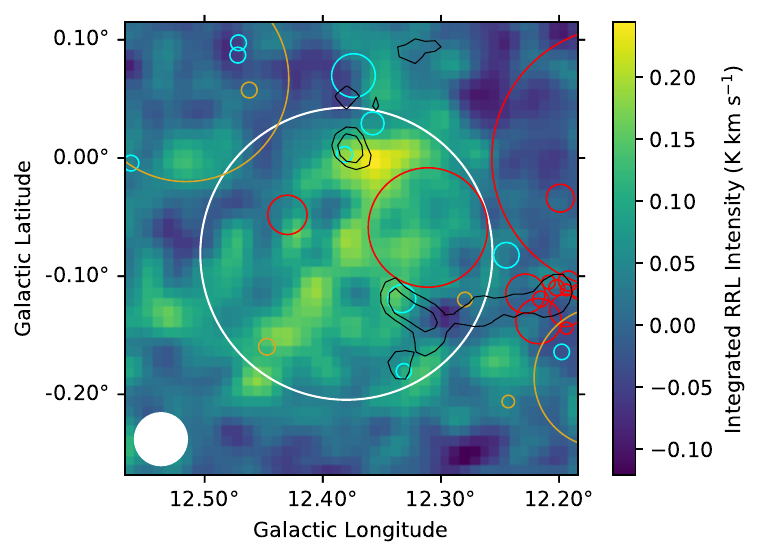}
\figsetplot{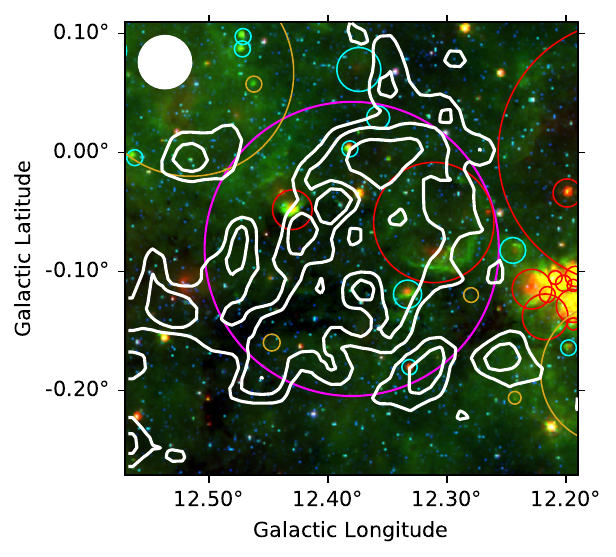}
\figsetgrpnote{The spectrum (1), RRL Moment~0 map with \cor\ contours (2), and IR image with RRL contours (3) of the AVF at galactic coordinates 012.380-00.081. The AVF has an approximate angular radius of 445\,\arcsec and an approximate  LSR velocity of 118.5\,\kms\. The GDIGS spectrum is Gaussian-smoothed with a FWHM of 5 \kms\. \cor\ contours on the RRL map are at 2$\sigma$ and 3.5$\sigma$ and are made with data spatially Gaussian-smoothed with a FWHM of 19\,\arcsec. RRL contours on the IR image are at $\sigma$, 1.8$\sigma$, and 3.1$\sigma$. In the IR image, 24\,\micron emission is represented in red, 8\,\micron emission is represented in green, and 3.6\,\micron emission is represented in blue. The solid white circles in the corners of the RRL Moment~0 map and the IR image indicate the size of the GDIGS beam.}
\figsetgrpend

\figsetgrpstart
\figsetgrpnum{5.8}
\figsetgrptitle{G013.782-00.152}
\figsetplot{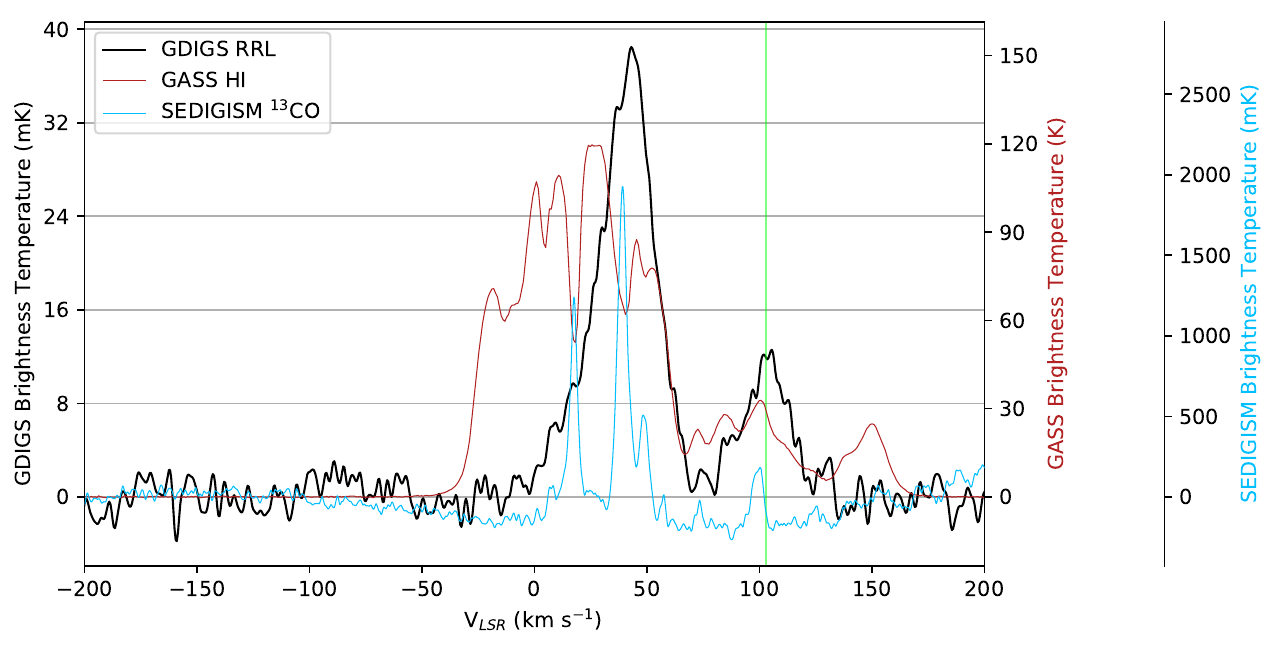}
\figsetplot{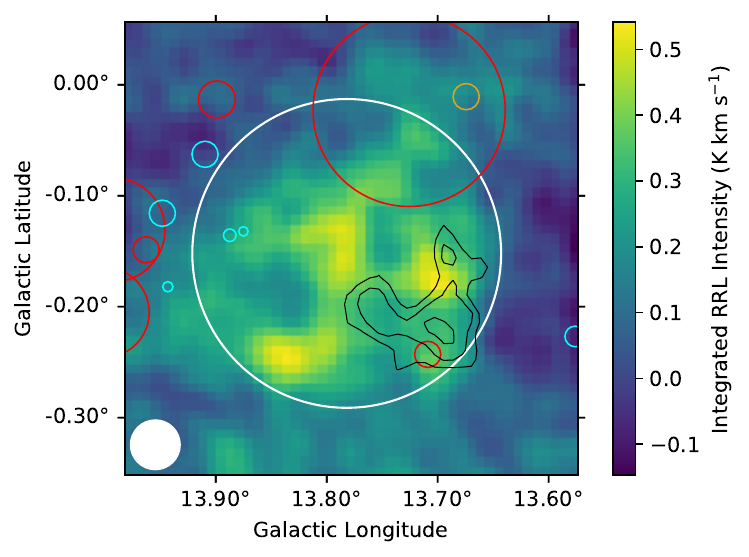}
\figsetplot{hv14_l13.782_n-0.152_v102.78_r501_percentile98_individualnorm_NIR_3color.pdf}
\figsetgrpnote{The spectrum (1), RRL Moment~0 map with \cor\ contours (2), and IR image with RRL contours (3) of the AVF at galactic coordinates 013.782-00.152. The AVF has an approximate angular radius of 501\,\arcsec and an approximate  LSR velocity of 102.8\,\kms\. The GDIGS spectrum is Gaussian-smoothed with a FWHM of 5 \kms\. \cor\ contours on the RRL map are at 2$\sigma$, 3.5$\sigma$, and 6.1$\sigma$ and are made with data spatially Gaussian-smoothed with a FWHM of 19\,\arcsec. RRL contours on the IR image are at $\sigma$, 1.8$\sigma$, and 3.1$\sigma$. In the IR image, 24\,\micron emission is represented in red, 8\,\micron emission is represented in green, and 3.6\,\micron emission is represented in blue. The solid white circles in the corners of the RRL Moment~0 map and the IR image indicate the size of the GDIGS beam.}
\figsetgrpend

\figsetgrpstart
\figsetgrpnum{5.9}
\figsetgrptitle{G014.549-00.106}
\figsetplot{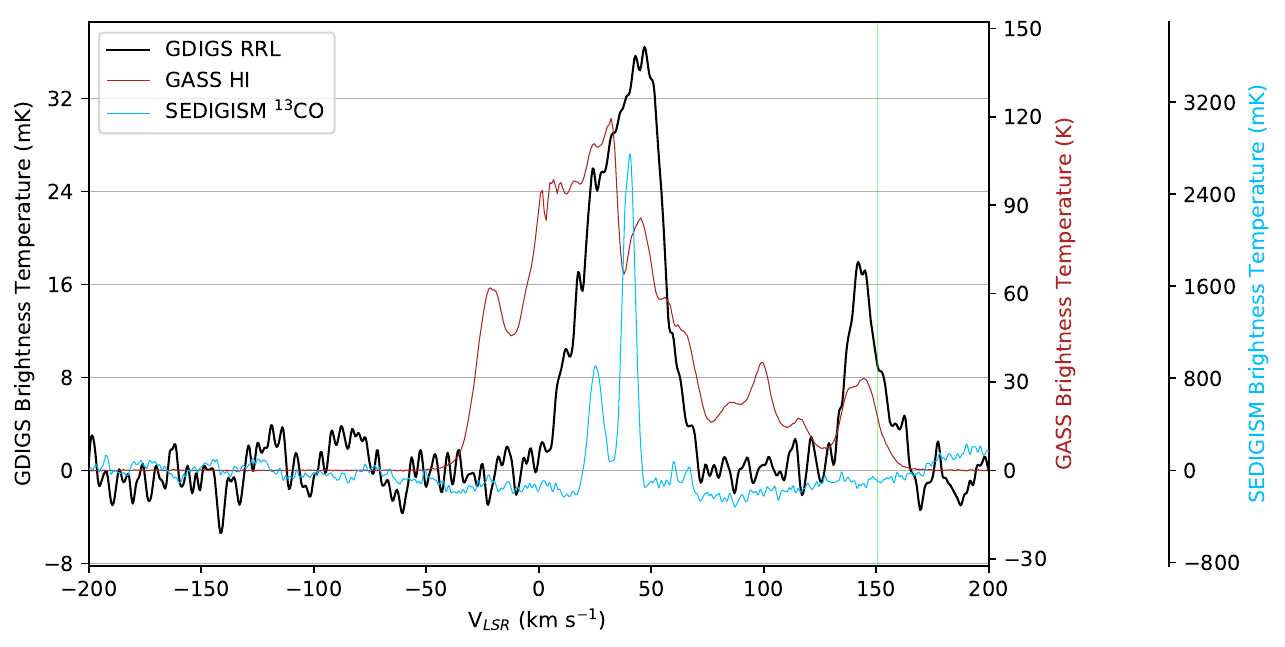}
\figsetplot{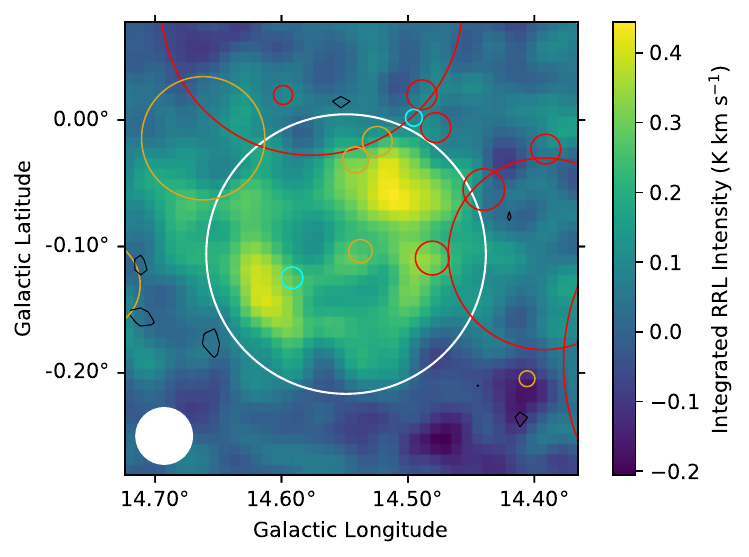}
\figsetplot{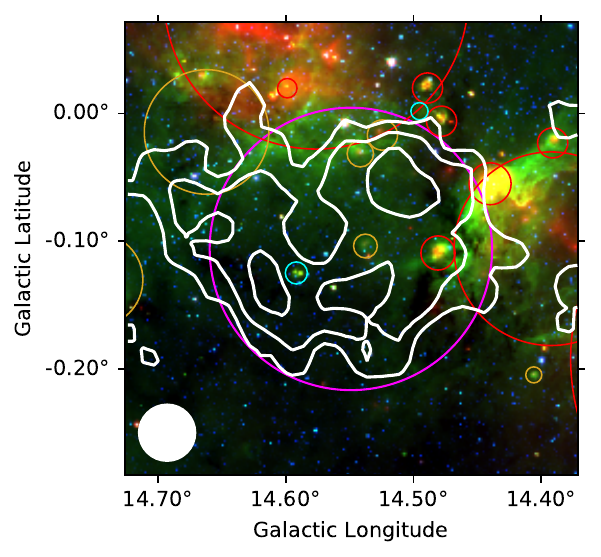}
\figsetgrpnote{The spectrum (1), RRL Moment~0 map with \cor\ contours (2), and IR image with RRL contours (3) of the AVF at galactic coordinates 014.549-00.106. The AVF has an approximate angular radius of 398\,\arcsec and an approximate  LSR velocity of 150.5\,\kms\. The GDIGS spectrum is Gaussian-smoothed with a FWHM of 5 \kms\. \cor\ contours on the RRL map are at 2$\sigma$ and are made with data spatially Gaussian-smoothed with a FWHM of 19\,\arcsec. RRL contours on the IR image are at $\sigma$, 1.8$\sigma$, and 3.1$\sigma$. In the IR image, 24\,\micron emission is represented in red, 8\,\micron emission is represented in green, and 3.6\,\micron emission is represented in blue. The solid white circles in the corners of the RRL Moment~0 map and the IR image indicate the size of the GDIGS beam.}
\figsetgrpend

\figsetgrpstart
\figsetgrpnum{5.10}
\figsetgrptitle{G020.206+00.036}
\figsetplot{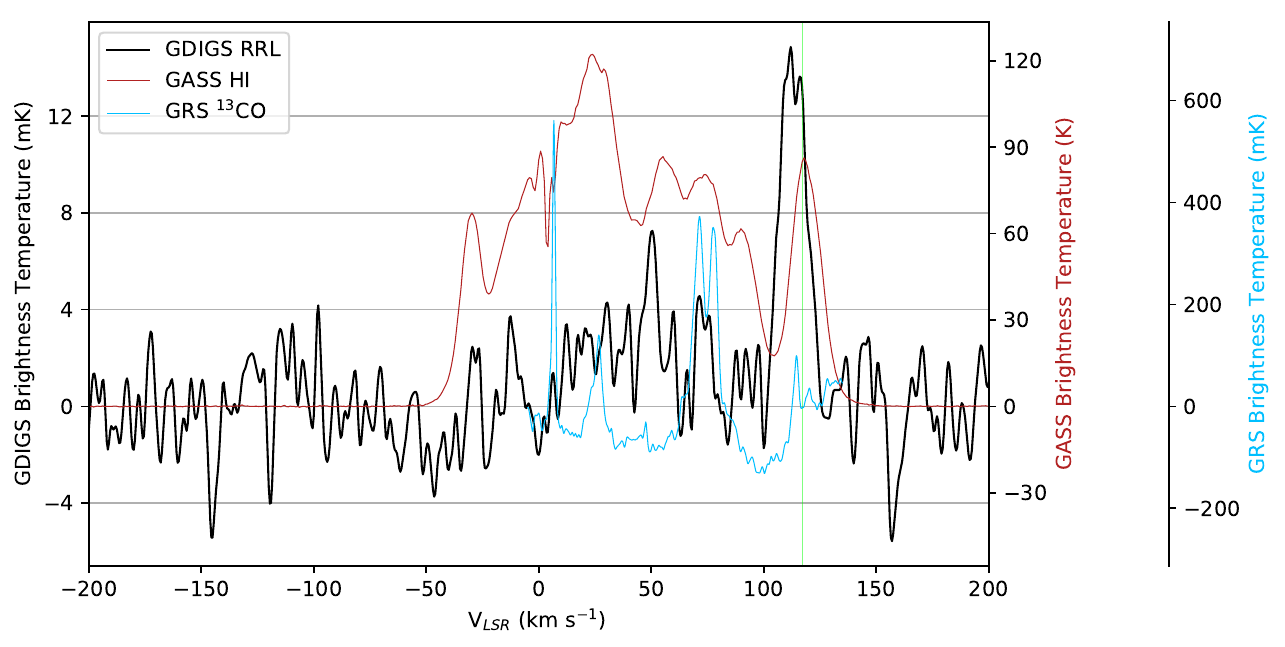}
\figsetplot{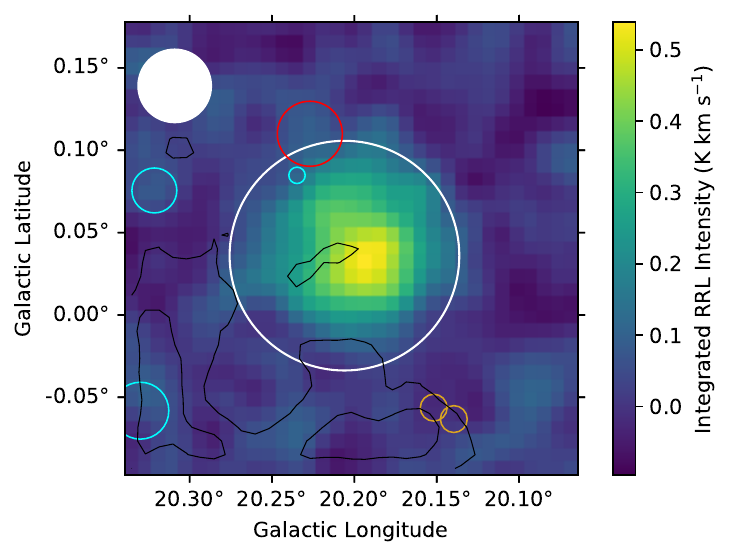}
\figsetplot{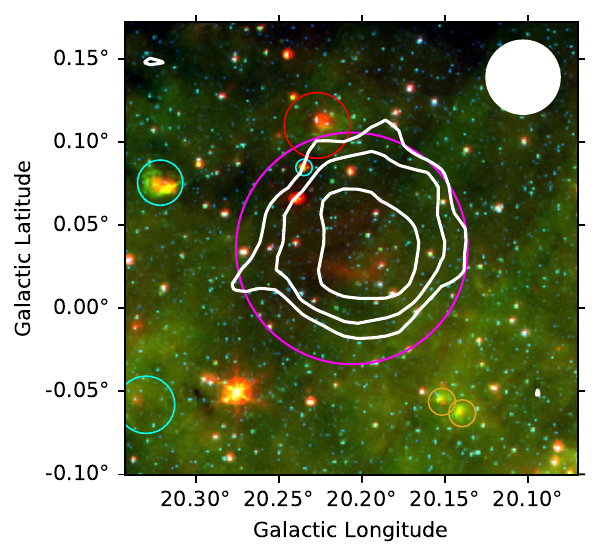}
\figsetgrpnote{The spectrum (1), RRL Moment~0 map with \cor\ contours (2), and IR image with RRL contours (3) of the AVF at galactic coordinates 020.206+00.036. The AVF has an approximate angular radius of 251\,\arcsec and an approximate  LSR velocity of 117.2\,\kms\. The GDIGS spectrum is Gaussian-smoothed with a FWHM of 5 \kms\. \cor\ contours on the RRL map are at 2$\sigma$ and 3.5$\sigma$ and are made with data spatially Gaussian-smoothed with a FWHM of 19\,\arcsec. RRL contours on the IR image are at $\sigma$, 1.8$\sigma$, and 3.1$\sigma$. In the IR image, 24\,\micron emission is represented in red, 8\,\micron emission is represented in green, and 3.6\,\micron emission is represented in blue. The solid white circles in the corners of the RRL Moment~0 map and the IR image indicate the size of the GDIGS beam.}
\figsetgrpend

\label{figset:anomaly_figs}

\figsetend

The WISE Catalog contains two known \hii regions that also have unusually high velocities: G002.614+00.133, which has a velocity of 102.4\,\kms \citep{lockman89}, and G007.700$-$00.123, which has a velocity of 151.5\,\kms \citep{anderson18b}. The AVFs  may be of a related class. Both of the known \hii\ regions, however, have coincident 24\,\micron\ emission, whereas the AVFs lack such emission. 
Roughly 80\% of \hii regions have associated \cor emission \citep{anderson09c}. Meanwhile, only 6 out of the 10 AVFs' spectra show \cor emission at the same velocity as the RRL emission. Only 1 out of the 10 AVFs, G355.612+00.314, shows \cor spatially associated with the RRL emission, and that AVF's RRL emission is not associated with MIR emission. Based on the lack of \cor and 24\,\micron\, emission from these objects, we consider it unlikely that they are \hii regions. To what class of object the AVFs belong remains undetermined.

The LV diagram of RRL emission presented in \citet{Hou22} shows some compact emission at velocities between 100 and 150\,\kms.  These features, however, have very narrow linewidths indicating that they are noise.

Below, we discuss noteworthy features of the most interesting individual AVFs.
\begin{itemize}
    \item AVF G355.612+00.314 is the only AVF observed in the fourth Galactic quadrant. It is presented as an example in Figure~\ref{fig:AVF_samples} because it demonstrates the aspects of AVFs outlined above: multiple bright areas connected by diffuse emission which completely surrounds them, spatially coincident \cor emission, 24\,\micron\ MIR emission associated with another object, spectrally coincident \hi and \cor emission, and several coincident but unassociated objects from the WISE catalog. 
    This AVF also falls within the velocity range of Bania's Clump 1, and sits on the edge of the Clump~1 longitude range \citep{Bania77}, as seen in Figure~\ref{fig:LV}. The AVF may be part of the Clump~1 complex, but according to SEDIGISM \cor $2-1$ data it is not coincident spatially or spectrally with the bulk of the Clump~1 \cor emission.
    \item AVF G001.702$-$00.156 is the closest in longitude to the Galactic Center and, at $+232.0$ \,\kms, has the highest LSR velocity of all the observed AVFs. Visual inspection of its position on the LV diagram (Figure~\ref{fig:LV}) suggests it may be associated with a Galactic Center structure, likely the nuclear disk. 
    The nuclear disk, also referred to as the nuclear ring or central molecular zone (CMZ),
    is a zone of high-velocity gas located within the inner $\sim\!0.5\,\kpc$ of the Milky Way \citep{Sormani19}.
    We base this inference on the behavior of the known \hii\ regions and the RRL observations near the Galactic center, which roughly form a slanted spike structure on the LV diagram representing the nuclear disk \citep{Fux99}. A line drawn to continue the apparent structure would pass near this AVF. 
    This AVF's velocity is outside the velocity range of the \cor\ data used here, so no \cor\ comparison is made.
    \item AVF G003.158$-$00.040 sits at the edge of the velocity range of Bania's Clump~2 and is within its longitude range \citep{Bania77}, as seen in Figure~\ref{fig:LV}. Like G355.612+0.314 and Clump~1, G003.158$-$0.040 straddles the edge of the box representing the longitude-velocity bounds of Clump~2, and we are likewise unable to conclude whether or not this AVF is associated with the Clump.
    \item AVF G009.402+00.180 has an exceptionally small FWHM, at only 5.1\,\kms. This value implies a low plasma temperature, an unusually low amount of turbulence, or both. 
    \item AVF G013.782$-$00.152 is the AVF with the lowest LSR velocity in the first Galactic quadrant, and the second lowest LSR velocity overall after G355.612+00.314. 
    One velocity component of the WISE \hii region G013.709$-$00.243, at 97.8\,\kms, is within 10\,\kms of the AVF's velocity, but the region is small in size relative to the AVF and it is far from the AVF's centroid, so we cannot draw any conclusions in regard to a possible association.
    \item AVF G020.206+00.036 has the greatest Galactic longitude by more than 5\degree.  
    It is the closest in velocity to the bulk of the \hii emission, but is distinct enough to be noticeable on Figure \ref{fig:LV}. This AVF has a small amount of coincident 24\,\micron\ emission, but the morphology does not resemble that expected of an \hii\ region.
\end{itemize}

\subsection{FWHM Analysis}
The FWHM line width is related to the thermal and turbulent motions in the plasma.  We can compare the trends in the FWHM values between known WISE Catalog objects and newly discovered regions to check for additional differences between the categories.
To that end, we calculate the mean and standard deviation of the FWHM line widths for known WISE Catalog regions, including those with first spectroscopic detections reported in this paper, and for each category of newly discovered region. 
We exclude regions with multiple velocity components. These statistics, as well as the number of included sources in each category, are reported in Table~\ref{tab:fwhm_stats}. There are too few new \hii\ regions and AVFs to reasonably draw conclusions from their statistics. The mean FWHM of the DEZs is greater than that of the WISE Catalog objects, but is within one standard deviation. We therefore cannot conclude any additional distinctions between the DEZs and WISE Catalog regions from FWHM trends alone. 

\begin{deluxetable}{lccr}
\tabletypesize{\scriptsize}
\tablecaption{FWHM Line Width Statistics}
\label{tab:fwhm_stats}
\tablewidth{0pt}
\tablehead{
\colhead{Source Category} &
\colhead{Mean} &
\colhead{St. Dev.} &
\colhead{Number}
\\
\colhead{} &
\colhead{(\kms)} &
\colhead{(\kms)} &
\colhead{}
}
\startdata
WISE Catalog Objects & 25.1 & 7.2 & 2279 \\
New \hii\ Regions & 31.7 & 5.5 & 4 \\
DEZs & 32.2 & 11.5 & 21 \\
AVFs & 21.5 & 7.3 & 10
\enddata
\end{deluxetable}

\section{Summary}
We use GDIGS RRL data to identify and characterize RRL features from discrete sources of emission. We summarize these results below.
\begin{itemize}
    \item We identify from the GDIGS data the \hii\ region velocity of 35 sources with multiple previously-reported velocity components.  We hypothesize that for \hii regions with multiple observed velocities, the velocity or velocities not associated with the \hii region itself belong to DIG along the line of sight. 
    \item We detect RRL emission for 40 ``group'' \hii regions,
    27 \hii region candidates, and
    21 radio quiet \hii regions, all from the WISE catalog. In total, we move 89 WISE Catalog objects into the ``known'' \hii\ region category. 
    \item We detect RRL emission from eight locations that we believe to be previously unknown Galactic \hii regions.
    \item We detect 30 GDIGS discrete sources that lack coincident MIR emission, and therefore we conclude they are not \hii regions. We detect and characterize via Gaussian fitting the RRL emission from these locations. 
    \item Finally, we identify 10 locations of anomalously high-velocity RRL emission, which we refer to as AVFs.  We examine the emission at these locations using \cor, \hi, and MIR data. Based on this analysis we conclude that the AVFs are likely not \hii regions. Beyond that, we do not have sufficient information to determine to what class of object they belong.
\end{itemize}

We present Gaussian fit parameters for 96 \hii regions, 30 DEZs, and 10 anomalous velocity features. Of these, the 30 DEZs, the 10 AVFs, and eight of the \hii regions were previously unknown; the rest lacked observed RRL spectra. These parameters, specifically the LSR velocities, allow calculation of the kinematic distance to these objects.  
The kinematic distance in turn allows determination of other physical parameters including physical extent of the regions.

\begin{appendix}

\section{WISE Catalog}
We have updated the WISE Catalog of Galactic \hii\ Regions website\footnote{http://astro.phys.wvu.edu/wise} with results from these observations. This includes the addition of LSR velocities for 88 \hii regions; updated velocities for 39 \hii regions with multiple velocity components; and 8 new \hii regions.

\end{appendix}

%% %% %% %% %% 

%%%%%%%%%%%%%%%%%%%%%%%%%%%%%%%%%%%%%%%%%%%%%%%%%%
\begin{acknowledgments}
This work is supported by NSF grant AST1516021 to LDA.
\nraoblurb\
The Green Bank Observatory is a facility of the National Science Foundation operated under cooperative agreement by Associated Universities, Inc.
We thank the staff at the Green Bank Observatory for their hospitality and friendship during the observations and data
reduction. We thank West Virginia University for its financial support of GBT operations, which enabled some of the observations for this project.

Part of this publication is based on data acquired with the Atacama Pathfinder Experiment (APEX) under programs 092.F-9315 and 193.C-0584. APEX is a collaboration among the Max-Planck-Institut fur Radioastronomie, the European Southern Observatory, and the Onsala Space Observatory. The processed data products are available from the SEDIGISM survey database located at https://sedigism.mpifr-bonn.mpg.de/index.html, which was constructed by James Urquhart and hosted by the Max Planck Institute for Radio Astronomy.

This publication makes use of molecular line data from the Boston University-FCRAO Galactic Ring Survey (GRS). The GRS is a joint project of Boston University and Five College Radio Astronomy Observatory, funded by the National Science Foundation under grants AST-9800334, AST-0098562, AST-0100793, AST-0228993, \& AST-0507657.

T.V.W. is supported by a National Science Foundation Astronomy and Astrophysics Postdoctoral Fellowship under award AST-2202340.

\end{acknowledgments}

%%%%%%%%%%%%%%%%%%%%%%%%%%%%%%%%%%%%%%%%%%%%%%%%%%%%%%%%%%%%%%%%%%%%%%%%%%%%%%%%

\software{AstroPy \citep{astropy2013, astropy2018, astropy2022}}

\bibliographystyle{aasjournal}
\bibliography{ref.bib}

\end{document}

%% file: preamble.tex
\usepackage{xspace}
\usepackage{graphicx}
\usepackage{natbib}
\usepackage{amsmath}
\usepackage{booktabs}
\usepackage{multirow}
\usepackage{afterpage}
\usepackage[counterclockwise, figuresleft]{rotating}
\newcommand{\nraoblurb}{The National Radio Astronomy Observatory is
a facility of the National Science Foundation operated under cooperative
agreement by Associated Universities, Inc.}
\newcommand{\hide}[1]{}

\newcommand{\y}[1]{\ensuremath{y_{#1}}\xspace}

\newcommand{\gl}{\ensuremath{\ell}\xspace}
\newcommand{\gb}{\ensuremath{{\it b}}\xspace}
\newcommand{\absb}{\ensuremath{\vert\,\gb\,\vert}\xspace}

\newcommand{\kms}{\ensuremath{\,{\rm km\,s^{-1}}}\xspace}

\newcommand{\m}{\ensuremath{\,{\rm m}}\xspace}
\newcommand{\cm}{\ensuremath{\,{\rm cm}}\xspace}

\newcommand{\kpc}{\ensuremath{\,{\rm kpc}}\xspace}
\newcommand{\K}{\ensuremath{\,{\rm K}}\xspace}
\newcommand{\mK}{\ensuremath{\,{\rm mK}}\xspace}

\newcommand{\myr}{\ensuremath{\,{\rm Myr}}\xspace}

\newcommand{\ghz}{\ensuremath{\,{\rm GHz}}\xspace}
\newcommand{\percc}{\ensuremath{\,{\rm cm^{-3}}}\xspace}

\newcommand{\degree}{\ensuremath{^\circ}\xspace}

\newcommand{\arcmper}{\ensuremath{\rlap.{^{\prime}}}}

\newcommand{\jyb}{\ensuremath{{\rm \,Jy\,beam^{-1}}}\xspace}

\newcommand{\h}[1]{\ensuremath{^{#1}{\rm H}}\xspace}
\newcommand{\hi}{{\rm H\,{\footnotesize I}}\xspace}
\newcommand{\hii}{{\rm H\,{\footnotesize II}}\xspace}

\newcommand{\hna}{\ensuremath{{\rm H}{\rm n}\alpha\xspace}}

\newcommand{\co} {\ensuremath{^{\rm 12}{\rm CO}}\xspace}
\newcommand{\cor}{\ensuremath{^{\rm 13}{\rm CO}}\xspace}

\RequirePackage{rotating}

%% file: gdigs_multvel5.tab
G355.739+00.131		&	10.7	&	$-$77.4	\\
G355.803$-$00.274	&	$-$31.5	&	3.1	\\
G359.386$-$00.082	&	$-$134.2&	$-$4.8	\\
G359.648$-$00.083	&	$-$43.8	&	0.8	\\
G359.717$-$00.036	&	$-$7.2	&	$-$60.0	\\
G359.727$-$00.037	&	$-$7.2	&	$-$60.0	\\
G000.729$-$00.103	&	83.2	&	105.3	\\
G007.268$-$00.060	&	47.0	&	17.5	\\
G007.380$-$00.264	&	31.8	&	9.7	\\
G010.237$-$00.080	&	12.0	&	51.2	\\
%G010.321$-$00.266	&	33.4	&	0.8	\\
G010.331$-$00.278	&	33.4	&	0.8	\\
G010.585$-$00.051	&	65.8	&	42.2	\\
G011.840$-$00.191	&	49.7	&	9.0	\\
G011.888$-$00.224	&	4.0	&	46.5	\\
G012.311$-$00.059	&	67.3	&	30.5	\\
G012.726$-$00.219	&	34.9	&	64.4	\\
G015.010+00.049		&	26.5	&	63.5	\\
G018.914$-$00.329	&	68.0	&	36.2	\\
G019.466+00.168		&	19.9	&	70.8; 109.3	\\
G019.494$-$00.150	&	32.1	&	54.6	\\
G022.790$-$00.347	&	112.1	&	69.3	\\
G022.987$-$00.155	&	100.2	&	76.6	\\
G023.257$-$00.453	&	61.6	&	89.0	\\
G024.826$-$00.265	&	44.7	&	109.8; 120.8	\\
G024.919+00.294		&	108.6	&	135.1	\\
G025.183+00.117		&	99.6	&	47.7;56.7	\\
G025.599$-$00.252	&	122.3	&	64.3; 114.2	\\
G026.306+00.073		&	92.4	&	35.8	\\
G028.605$-$00.013	&	41.9	&	92.8	\\
G030.212$-$00.157	&	98.6	&	66.2	\\
G030.248$-$00.102	&	102.9	&	67.9	\\
G030.442$-$00.004	&	38.9	&	92.7; 109.2	\\
G030.478$-$00.091	&	102.1	&	78.7; 47.2	\\
%G030.533+00.021		&	46.1	&	95.2	\\
%G030.545+00.027		&	46.1	&	95.2	\\
%G030.547+00.014		&	46.1	&	95.2	\\
G030.613$-$00.617	&	12.2	&	87.8; 110.2	\\
G030.867+00.008		&	96.1	&	44.8	

%% file: gdigs_groups9.tab
% Name           & glong & glat     & T_L  &sigTL& VLSR &sigV & FWHM &sigDV& rms & S/N\\
G004.613$-$00.150 & 4.614 & $-$0.145 & 33.3 & 1.5 & 18.4 & 0.8 & 34.4 & 1.8 & 3.3 & 27.0\\
G008.153$-$00.198 & 8.098 & $-$0.093 & 25.4 & 0.3 & 40.9 & 0.2 & 31.4 & 0.5 & 1.1 & 59.1 \\
G008.181$+$00.321 & 8.183 & 0.320 & 33.5 & 1.0 & 22.9 & 0.4 & 23.9 & 0.8 & 2.0 & 37.4\\
G008.472$-$00.263 & 8.472 & $-$0.262 & 36.3 & 1.4 & 36.0 & 0.6 & 32.8 & 1.5 & 3.4 & 27.9 \\
G008.474$-$00.281 & 8.476 & $-$0.280 & 69.0 & 1.7 & 38.6 & 0.3 & 24.8 & 0.7 & 3.2 & 49.0 \\
G009.381$-$00.067 & 9.385 & $-$0.076 & 9.1 & 0.9 & 49.1 & 2.6 & 55.2 & 6.1 & 2.6 & 11.9 \\
G009.599$+$00.165 & 9.597 & 0.168 & 40.2 & 2.0 & 4.6 & 0.5 & 20.7 & 1.2 & 3.6 & 23.2 \\
G010.227$-$00.346 & 10.233 & $-$0.349 & 99.5 & 1.0 & 15.4 & 0.2 & 36.8 & 0.4 & 2.4 & 114.8 \\
G010.321$-$00.266 & 10.322 & $-$0.265 & 85.2 & 1.5 & 32.0 & 0.3 & 30.0 & 0.6 & 2.9 & 73.4 \\
G010.321$-$00.266* & 10.322 & $-$0.265 & 43.0 & 0.6 & 1.5 & 1.2 & 19.4 & 0.4 & 2.9 & 29.8 \\
G010.441$+$00.012a & 10.441 & 0.011 & 104.1 & 1.6 & 71.9 & 0.2 & 22.6 & 0.4 & 2.7 & 83.7 \\
G010.441$+$00.012b & 10.441 & 0.011 & 18.2 & 1.8 & 42.9 & 6.1 & 54.8 & 9.0 & 2.7 & 22.8\\
G010.453$-$00.006a & 10.455 & $-$0.005 & 62.6 & 1.5 & 69.1 & 0.3 & 23.8 & 0.7 & 2.4 & 58.1 \\
G010.453$-$00.006b & 10.455 & $-$0.005 & 29.7 & 0.6 & 42.3 & 0.7 & 30.1 & 1.5 & 2.4 & 31.0 \\
G010.497$+$00.028a & 10.498 & 0.029 & 42.9 & 0.8 & 66.6 & 0.2 & 24.0 & 0.6 & 2.9 & 33.1 \\
G010.497$+$00.028b & 10.498 & 0.029 & 27.0 & 0.4 & 44.5 & 0.3 & 21.0 & 0.8 & 2.9 & 19.5 \\
%G010.574$-$00.412 & 10.587 & $-$0.414 & 165.6 & 1.4 & $-$1.0 & 0.1 & 21.4 & 0.2 & 3.0 \\
G010.617$-$00.405 & 10.619 & $-$0.400 & 287.8 & 2.0 & $-$1.4 & 0.1 & 24.6 & 0.2 & 5.1 & 127.8 \\
G010.966$+$00.022 & 10.968 & 0.024 & 66.2 & 1.8 & 17.4 & 0.4 & 29.1 & 0.9 & 4.0 & 39.8 \\
G012.683$-$00.185a & 12.682 & $-$0.184 & 226.7 & 2.2 & 52.8 & 0.1 & 24.5 & 0.4 & 4.7 & 109.0 \\
G012.683$-$00.185b & 12.682 & $-$0.184 & 133.3 & 2.5 & 34.3 & 0.3 & 18.7 & 0.8 & 4.7 & 56.0 \\
G012.692$-$00.251 & 12.688 & $-$0.245 & 197.7 & 1.9 & 33.5 & 0.1 & 16.1 & 0.2 & 3.3 & 109.4 \\
G012.692$-$00.251* & 12.688 & $-$0.245 & 105.2 & 1.6 & 55.1 & 0.2 & 21.0 & 0.5 & 3.3 & 66.7\\
G012.820$-$00.238a & 12.873 & $-$0.234 & 161.9 & 1.5 & 32.6 & 0.1 & 25.5 & 0.3 & 2.7 & 138.3 \\
G012.820$-$00.238b & 12.873 & $-$0.234 & 25.4 & 1.8 & 63.1 & 1.9 & 18.7 & 3.4 & 2.7 & 18.6 \\
G012.884$-$00.237 & 12.818 & $-$0.243 & 326.0 & 2.0 & 32.9 & 0.1 & 27.6 & 0.2 & 3.8 & 205.8 \\
G012.884$-$00.237* & 12.818 & $-$0.243 & 55.0 & 2.2 & 59.2 & 1.2 & 22.7 & 2.3 & 3.9 & 31.5 \\
G012.891$-$00.195a & 12.894 & $-$0.195 & 58.7 & 0.6 & 33.7 & 0.1 & 25.0 & 0.3 & 3.5 & 38.3 \\
G012.891$-$00.195b & 12.894 & $-$0.195 & 21.0 & 0.9 & 62.6 & 0.5 & 22.3 & 1.4 & 3.5 & 12.9 \\
G016.258$-$00.091 & 16.254 & $-$0.090 & 34.9 & 0.9 & 42.8 & 0.4 & 34.6 & 1.0 & 2.0 & 46.9 \\
G016.352$-$00.179 & 16.349 & $-$0.176 & 96.5 & 1.5 & 45.4 & 0.2 & 26.6 & 0.5 & 3.4 & 66.8 \\
G017.073$-$00.096 & 17.082 & $-$0.092 & 47.8 & 0.5 & 93.7 & 0.1 & 24.6 & 0.3 & 2.6 & 41.6 \\
G017.073$-$00.096* & 17.082 & $-$0.092 & 10.0 & 0.3 & 36.7 & 0.9 & 41.5 & 2.5 & 2.6 & 11.3 \\
G018.441$+$00.013 & 18.441 & 0.014 & 46.5 & 1.8 & 61.9 & 0.6 & 33.2 & 1.5 & 3.8 & 32.2 \\
G024.093$+$00.455 & 24.094 & 0.456 & 15.0 & 0.2 & 85.9 & 0.3 & 44.1 & 0.8 & 1.7 & 26.8 \\
G024.525$-$00.041 & 24.528 & $-$0.046 & 60.2 & 0.4 & 97.8 & 0.2 & 54.0 & 0.5 & 1.8 & 112.2 \\
G024.640$-$00.142a & 24.648 & $-$0.139 & 100.2 & 0.5 & 103.3 & 0.1 & 41.9 & 0.3 & 2.6 & 53.5 \\
G024.640$-$00.142b & 24.648 & $-$0.139 & 47.5 & 0.5 & 38.8 & 0.2 & 41.2 & 0.6 & 2.6 & 113.9 \\
G025.179$+$00.038 & 25.177 & 0.028 & 17.3 & 0.3 & 44.4 & 0.4 & 48.2 & 1.1 & 1.6 & 34.3 \\
G025.179$+$00.038* & 25.177 & 0.028 & 12.0 & 0.2 & 91.3 & 0.7 & 56.0 & 1.9 & 1.6 & 25.6 \\
G028.185$+$00.314 & 28.186 & 0.316 & 7.3 & 0.8 & 45.8 & 2.4 & 40.0 & 6.2 & 1.9 & 11.1 \\
G028.185$+$00.314* & 28.186 & 0.316 & 8.7 & 0.8 & 95.1 & 2.3 & 46.0 & 6.1 & 1.9 & 14.2 \\
G028.670$+$00.011 & 28.670 & 0.008 & 122.0 & 0.9 & 102.5 & 0.1 & 28.6 & 0.2 & 4.1 & 72.7 \\
G028.670$+$00.011* & 28.670 & 0.008 & 15.8 & 0.6 & 41.1 & 0.3 & 19.1 & 0.9 & 4.1 & 7.7 \\
G028.702$+$00.014a & 28.704 & 0.014 & 91.6 & 0.9 & 103.7 & 0.2 & 33.3 & 0.4 & 3.7 & 65.2 \\
G028.702$+$00.014b & 28.704 & 0.014 & 15.2 & 0.5 & 39.3 & 0.4 & 23.7 & 1.0 & 3.7 & 9.1 \\
G029.814$-$00.069 & 29.862 & $-$0.083 & 204.6 & 1.0 & 97.8 & 0.1 & 24.1 & 0.1 & 1.6 & 286.7 \\
G029.862$-$00.059 & 29.862 & $-$0.059 & 286.2 & 1.9 & 96.4 & 0.1 & 21.5 & 0.2 & 3.3 & 183.6 \\
G029.889$-$00.007 & 29.887 & $-$0.002 & 220.5 & 1.0 & 96.7 & 0.1 & 26.6 & 0.1 & 1.6 & 324.6 \\
G029.901$-$00.029 & 29.904 & $-$0.025 & 147.7 & 0.7 & 96.4 & 0.1 & 26.4 & 0.1 & 1.2 & 288.8 \\
G029.928$-$00.057 & 29.927 & $-$0.058 & 464.2 & 1.7 & 98.5 & 0.1 & 29.4 & 0.1 & 2.8 & 410.5 \\
G029.955$-$00.093 & 29.963 & $-$0.091 & 173.1 & 1.1 & 96.5 & 0.1 & 28.6 & 0.2 & 2.1 & 201.3 \\
G030.007$-$00.084 & 30.010 & $-$0.085 & 155.8 & 1.2 & 95.6 & 0.1 & 23.9 & 0.2 & 2.5 & 139.1 \\
G031.264$+$00.031a & 31.265 & 0.033 & 62.9 & 0.6 & 101.7 & 0.1 & 26.0 & 0.3 & 2.6 & 56.3 \\
G031.264$+$00.031b & 31.265 & 0.033 & 18.6 & 0.3 & 35.9 & 0.4 & 45.7 & 1.1 & 2.6 & 22.1 \\
G031.309$+$00.044 & 31.310 & 0.039 & 63.3 & 1.7 & 101.8 & 0.1 & 23.7 & 0.3 & 2.1 & 67.1 \\
G031.309$+$00.044* & 31.310 & 0.039 & 12.8 & 0.2 & 38.3 & 0.4 & 41.7 & 1.2 & 2.1 & 18.0 \\
G031.422$-$00.256 & 31.432 & $-$0.249 & 106.4 & 1.1 & 88.5 & 0.2 & 32.1 & 0.4 & 1.8 & 152.9 \\
G031.422$-$00.256* & 31.432 & $-$0.249 & 14.8 & 0.2 & 41.5 & 0.5 & 37.5 & 1.2 & 1.8 & 23.0

%% file: gdigs_candidates11.tab
% Name		  & glong   &   glat   & T_L & sigTL & VLSR   & sigV & FWHM & sigDV & RMS & S/N\\
G355.681+00.143   & 355.681 & 0.147 & 40.8 & 0.5 & $-$76.5 & 0.1 & 14.3 & 0.2 & 1.8 & 39.1 \\
G355.681+00.143*  & 355.681 & 0.147 & 24.5 & 0.4 & 7.0     & 0.2 & 18.1 & 0.4 & 1.8 & 26.4 \\
G355.727$-$00.212 & 355.727 & 0.212 & 58.3 & 0.9 & $-$76.9 & 0.1 & 11.1 & 0.2 & 2.1 & 42.2 \\
G355.727$-$00.212* & 355.727 & 0.212 & 20.7 & 0.3 & 7.7    & 0.2 & 22.2 & 0.4 & 2.1 & 21.2 \\
G359.690+00.065   & 359.690 & 0.065 & 22.4 & 0.7 & $-$90.0 & 0.7 & 41.1 & 1.6 & 3.0 & 24.9 \\
G359.690+00.065*  & 359.690 & 0.065 & 27.9 & 0.7 & 5.4     & 0.3 & 22.5 & 0.7 & 3.0 & 22.1 \\
G000.132+00.039   & 0.132   & 0.040 & 341.2 & 1.3 & $-$17.3 & 0.1 & 57.2 & 0.3 & 12.4 & 95.0 \\
%G000.132+00.039b & 0.132 & 0.040 & 68.0  & 3.0 & 24.3  & 2.5 & 62.5 & 3.1 & 10.3 \\
G000.551+00.060   & 0.551   & 0.060 & 60.4 & 0.6 & 52.0    & 0.3 & 50.2 & 0.7 & 3.1 & 63.0 \\
G000.551+00.060*  & 0.551   & 0.060 & 22.2 & 0.5 & 8.7     & 1.3 & 47.0 & 3.2 & 3.1 & 22.4 \\
G000.591+00.050   & 0.592   & 0.050 & 58.1 & 0.8 & 56.0    & 0.3 & 37.7 & 0.7 & 3.6 & 45.2 \\
G000.591+00.050*  & 0.592   & 0.050 & 21.3 & 0.5 & 6.2     & 0.6 & 42.2 & 1.5 & 3.6 & 17.6 \\
G016.770+00.019   & 16.770  & 0.020 & 18.8 & 0.9 & 32.8    & 0.8 & 32.3 & 1.9 & 5.5 & 8.9 \\
G020.427$-$00.074 & 20.427  & $-$0.074 & 24.6 & 0.6 & 49.2 & 0.4 & 28.2 & 0.8 & 3.8 & 15.7\\
G021.404$-$00.238 & 21.404  & $-$0.237 & 16.5 & 0.8 & 80.1 & 0.7 & 30.5 & 1.6 & 4.4 & 9.5 \\
G021.474$-$00.156 & 21.475  & $-$0.156 & 8.6  & 0.4 & 77.8 & 1.0 & 49.3 & 2.5 & 2.6 & 10.6 \\
G021.494$-$00.159 & 21.494  & $-$0.158 & 8.6  & 0.3 & 80.4 & 1.0 & 53.2 & 2.4 & 2.6 & 11.0 \\
G021.521$-$00.253 & 21.521  & $-$0.253 & 11.5 & 0.2 & 72.8 & 0.6 & 62.6 & 1.5 & 2.1 & 19.8 \\
G021.827+00.042a & 21.827  & 0.043  & 21.2 & 0.4 & 22.4  & 0.3 & 39.2 & 0.9 & 2.0  & 30.3 \\
G021.827+00.042b & 21.827  & 0.043  & 8.6  & 0.2 & 90.4  & 0.7 & 44.0 & 2.0 & 2.0  & 13.0 \\
G024.082+00.121 & 24.082  & 0.121  & 18.9  & 0.1 & 88.6  & 0.3 & 70.2 & 0.7 & 1.3  & 55.6 \\
%G024.082+00.121* & 24.082  & 0.121  & 18.7  & 0.2 & 92.6 & 0.7 & 62.1 & 1.6 & 1.3  \\
G024.114$-$00.183 & 24.114  & $-$0.183  & 21.2  & 0.1 & 80.2 & 0.2 & 49.7 & 0.5 & 1.1 & 62.0 \\
%G024.114$-$00.183* & 24.114  & $-$0.183  & 28.2  & 2.1 & 102.4 & 4.1 & 82.6 & 3.6 & 1.1  \\
G025.330$-$00.461 & 25.330  & $-$0.460 & 15.9 & 0.3 & 61.0 & 0.4 & 53.9 & 1.1 & 1.1 & 48.5 \\
%G025.330$-$00.461* & 25.330  & $-$0.460 & 22.5 & 2.1 & 12.4 & 8.6 & 129.8 & 8.0 & 1.1  \\
G026.613+00.309 & 26.614  & 0.309  & 17.0  & 0.4 & 106.5 & 0.4 & 28.2 & 0.9 & 1.7 & 24.2 \\
%G026.613+00.309b & 26.614  & 0.309  & 4.4   & 0.5 & 39.6  & 1.3 & 23.0 & 3.0 & 2.4  \\
G026.718$-$00.385  & 26.719  & $-$0.384 & 18.5  & 0.4 & 86.1  & 0.4 & 37.1 & 1.0 & 1.5 & 34.3 \\
G027.720$-$00.106  & 27.720  & $-$0.105 & 18.1  & 0.5 & 91.6  & 0.3 & 27.7 & 0.8 & 2.3 & 18.9 \\
G028.024$-$00.162  & 28.024  & $-$0.162 & 26.5  & 0.3 & 103.3 & 0.2 & 33.0 & 0.4 & 1.9 & 36.6 \\
G028.078+00.144  & 28.079  & 0.144  & 17.8  & 0.4 & 87.9  & 0.5 & 44.5 & 1.3 & 2.8 & 19.4 \\
G028.438+00.033  & 28.438  & 0.033  & 28.2  & 0.7 & 94.5  & 0.4 & 33.5 & 0.9 & 4.6 & 16.2 \\
G029.007$-$00.289  & 29.009  & $-$0.288 & 13.9  & 0.2 & 92.9  & 0.3 & 39.8 & 0.6 & 1.0 & 40.0 \\
G031.243+00.274 & 31.243  & 0.274 & 24.3  & 0.6 & 98.9  & 0.3 & 26.1 & 0.7 & 3.0 & 18.9 \\
G031.243+00.274* & 31.243  & 0.274 & 9.2  & 0.5 & 38.9  & 0.9 & 30.9 & 2.1 & 3.0 & 7.8 \\
G032.034$-$00.375 & 32.034  & $-$0.375 & 23.6  & 0.4 & 44.4  & 0.2 & 21.1 & 0.5 & 1.8 & 27.5 \\
G032.034$-$00.375* & 32.034  & $-$0.375 & 6.3  & 0.3 & 85.4  & 1.0 & 35.6 & 2.6 & 1.8 & 9.5 \\
G037.319+00.162a  & 37.319  & 0.162 & 7.4 & 1.3 & 26.6 & 3.2 & 36.6 & 8.4 & 2.7 & 11.4 \\
%G037.319+00.162 & 37.319 & 0.162 & 6.9 & 0.4 & 50.9 & 2.4 & 89.2 & 7.9 & 2.6 & 11.4 \\
G037.319+00.162b & 37.319 & 0.162 & 6.7 & 1.0 & 62.7 & 6.2 & 56.9 & 15.6 & 2.7 & 8.5 \\
G037.320+00.168a  & 37.321  & 0.169 & 7.4 & 1.1 & 67.6 & 5.2 & 56.0 & 13.6 & 2.9 & 10.3 \\
%G037.320+00.168 & 37.321 & 0.169 & 7.6 & 0.4 & 44.8 & 3.7 & 74.2 & 9.6 & 2.9 & 10.3 \\
G037.320+00.168b & 37.321 & 0.169 & 6.7 & 1.3 & 28.6 & 5.7 & 47.0 & 14.6 & 2.9 & 7.2

%% file: gdigs_candidatesrq7.tab
% Name            & glong   & glat     & T_L   &sigTL& VLSR  & sigV& FWHM &sigDV& RMS & S/N \\
G359.901$-$00.320 & 359.901 & $-$0.319 & 100.1 & 1.0 & 15.0  & 0.1 & 24.2 & 0.3 & 2.3 & 97.8 \\
G000.408$-$00.881  & 0.408  & $-$0.881 & 54.7  & 0.5 & 14.0  & 0.1 & 18.9 & 0.2 & 2.5 & 43.4 \\
G000.456$-$00.808  & 0.456  & $-$0.808 & 46.6  & 0.4 & 14.0  & 0.1 & 20.1 & 0.2 & 2.1 & 45.4 \\
G000.477$-$00.130  & 0.477  & $-$0.130 & 56.7  & 0.5 & 29.8  & 0.3 & 54.7 & 0.7 & 2.3 & 83.3 \\
G000.834$-$00.457  & 0.835  & $-$0.457 & 22.4  & 0.3 & 17.7  & 0.1 & 22.7 & 0.3 & 0.9 & 54.1 \\
G001.119$-$00.266  & 1.119  & $-$0.266 & 42.4  & 0.6 & 2.4   & 0.2 & 24.7 & 0.4 & 1.7 & 56.6 \\
G003.627+00.087    & 3.628  & 0.087    & 18.8  & 0.5 & 9.4   & 0.4 & 36.4 & 1.1 & 3.1 & 16.7 \\
G004.270+00.038    & 4.270  & 0.039    & 18.9  & 0.7 & 9.8   & 0.5 & 28.2 & 1.2 & 3.6 & 12.7 \\
G006.256$-$00.074  & 6.257  & $-$0.073 & 52.3  & 0.6 & 20.8  & 0.2 & 33.2 & 0.4 & 2.7 & 51.0 \\
G008.359+00.003    & 8.360  & 0.004    & 29.7  & 0.4 & 26.0  & 0.3 & 48.3 & 0.8 & 1.9 & 49.6 \\
G008.685$-$00.047  & 8.685  & $-$0.046 & 15.4  & 0.2 & 30.1  & 0.4 & 51.6 & 1.1 & 1.4 & 36.1 \\
G009.070+00.082    & 9.071  & 0.082    & 9.9   & 0.3 & 31.2  & 0.8 & 57.1 & 2.2 & 1.9 & 18.0 \\
G010.518$-$00.289  & 10.518 & $-$0.288 & 26.9  & 0.7 & 0.0   & 0.2 & 19.0 & 0.6 & 1.5 & 35.7 \\
G010.518$-$00.289* & 10.518 & $-$0.288 & 4.4   & 0.2 & 38.2  & 1.8 & 56.7 & 4.3 & 1.5 & 10.1 \\
G011.822$-$00.097  & 11.822 & $-$0.097 & 49.7  & 1.2 & 61.3  & 0.3 & 22.5 & 0.6 & 1.8 & 59.8 \\
G011.822$-$00.097* & 11.822 & $-$0.097 & 12.9  & 0.4 & 9.7   & 0.4 & 25.9 & 1.0 & 1.8 & 16.7 \\
G011.822$-$00.097* & 11.822 & $-$0.097 & 6.5   & 0.2 & 118.5 & 0.9 & 34.5 & 2.7 & 1.8 & 9.7 \\
G011.887$-$00.253a & 11.888 & $-$0.253 & 37.9  & 0.4 & 48.9  & 0.1 & 22.3 & 0.3 & 1.7 & 48.1 \\
G011.887$-$00.253b & 11.888 & $-$0.253 & 12.8  & 0.3 & 7.7   & 0.4 & 29.8 & 1.0 & 1.7 & 18.8 \\
G012.051$-$00.171  & 12.051 & $-$0.171 & 37.1  & 0.7 & 39.8  & 0.2 & 22.6 & 0.5 & 4.2 & 19.2 \\
G012.161$-$00.186  & 12.161 & $-$0.186 & 35.2  & 0.3 & 38.3  & 0.2 & 40.0 & 0.5 & 1.8 & 56.5 \\
G012.515+00.066    & 12.516 & 0.067    & 15.4  & 0.3 & 54.0  & 0.3 & 22.5 & 0.8 & 1.3 & 25.7 \\
G012.515+00.066*   & 12.516 & 0.067    & 34.4  & 0.4 & 28.8  & 0.1 & 24.4 & 0.4 & 1.3 & 59.7 \\
G014.068$-$00.432a & 14.068 & $-$0.432 & 31.1  & 0.4 & 26.0  & 0.2 & 32.8 & 0.6 & 1.8 & 45.2 \\
G014.068$-$00.432b & 14.068 & $-$0.432 & 18.1  & 0.2 & 47.9  & 0.6 & 40.5 & 1.3 & 1.8 & 29.2 \\
%G014.068$-$00.432c & 14.068 & $-$0.432 & 7.2  & 0.5 & 103.1 & 0.9 & 8.7  & 1.9 & 3.5 \\
G017.037+00.320    & 17.038 & 0.321    & 20.5  & 0.3 & 19.7  & 0.2 & 32.3 & 0.5 & 1.7 & 31.3 \\
G037.642+01.192    & 37.643 & 1.193    & 7.9   & 0.2 & 42.9  & 0.4 & 29.3 & 0.9 & 1.1 & 17.8

%% file: gdigs_newregions7.tab
% Name            & glong   & glat     & rad & T_L  &sigTL& VLSR  &sigV & FWHM &sigDV& RMS & S/N \\
G357.626$-$00.082 & 357.626 & $-$0.082 & 190 & 4.4  & 0.3 & 3.1   & 1.2 & 37.5 & 2.9 & 2.3 & 5.3 \\
G012.522$-$00.121 & 12.522  & $-$0.121 & 273 & 32.9 & 0.5 & 47.1  & 0.3 & 21.8 & 0.6 & 1.8 & 39.0 \\
G012.522$-$00.121* & 12.522 & $-$0.121 & 273 & 72.5 & 1.2 & 29.3  & 0.1 & 16.7 & 0.4 & 1.8 & 75.2 \\
G012.576+00.221   & 12.576  & 0.221    & 217 & 30.2 & 0.4 & 20.8  & 0.2 & 23.6 & 0.4 & 1.8 & 37.3 \\
G020.805+00.019   & 20.805  & 0.019    & 581 & 49.8 & 0.5 & 52.3  & 0.1 & 29.5 & 0.3 & 1.3 & 95.0 \\
%G024.147$-$00.547 & 24.147  & $-$0.547 & 179 & 19.2 & 0.5 & 101.9 & 0.4 & 28.7 & 1.1 & 4.1 \\
G028.567+00.002   & 28.567  & 0.002    & 191 & 68.4 & 0.5 & 97.5  & 0.1 & 33.5 & 0.3 & 2.7 & 67.0 \\
G028.567+00.002*  & 28.567  & 0.002    & 191 & 18.5 & 0.4 & 41.3  & 0.3 & 27.2 & 0.7 & 2.7 & 16.2 \\
G028.618+00.177a  & 28.618  & 0.178    & 655 & 21.4 & 0.4 & 103.0 & 0.3 & 31.1 & 0.6 & 1.0 & 54.5 \\
G028.618+00.177b  & 28.618  & 0.178    & 655 & 5.7  & 0.2 & 33.4  & 0.5 & 26.1 & 1.5 & 1.0 & 13.3 \\
G028.618+00.177c  & 28.618  & 0.178    & 655 & 5.3  & 0.2 & 68.2  & 0.9 & 33.4 & 3.4 & 1.0 & 14.0 \\
G030.820$-$00.764 & 30.820  & $-$0.764 & 666 & 18.0 & 0.2 & 81.4  & 0.2 & 36.1 & 0.6 & 0.9 & 54.9 \\
%G030.820$-$00.764* & 30.820 & $-$0.764 & 666 & 29.3 & 5.5 & 38.6  & 12.3 & 105.1 & 10.1 & 0.9 \\
G042.634+00.125   & 42.634  & 0.125    & 132 & 13.0 & 0.3 & 12.9  & 0.4 & 32.6 & 1.2 & 1.9 & 28.2 \\
G042.634+00.125*  & 42.634  & 0.125    & 132 & 8.7  & 0.3 & 63.1  & 0.6 & 36.4 & 1.5 & 1.9 & 20.0

%% file: gdigs_newregionsdig15.tab
% Name             & glong   & glat     & rad  & T_L  &sigTL& VLSR     &sigV & FWHM  &sigDV & rms & S/N \\
%G355.837+00.177    & 355.837 & 0.177    & 2880 & 20.6 & 0.3 & 5.6      & 0.2 & 23.1  & 0.4  & 0.6 & 73.4 \\
%G355.837+00.177*   & 355.837 & 0.177    & 2880 & 6.9  & 0.1 & $-$77.2  & 0.1 & 15.9  & 0.3  & 0.6 & 20.9 \\
G357.627$-$00.415  & 357.627 & $-$0.415 & 466  & 16.9 & 0.4 & 9.1      & 0.3 & 26.3  & 0.7  & 2.1 & 18.8 \\
%G358.170+00.243 & 358.170 & 0.243 & 190 & \nodata & \nodata & \nodata & \nodata & \nodata & \nodata & 6.1 \\
%G358.214+00.331 & 358.214 & 0.331 & 190 & \nodata & \nodata & \nodata & \nodata & \nodata & \nodata & 6.4 \\
%G358.336+00.292 & 358.336 & 0.292 & 190 & \nodata & \nodata & \nodata & \nodata & \nodata & \nodata & 5.4 \\
G358.414$-$00.122a & 358.414 & $-$0.122 & 229  & 4.1  & 0.4 & 12.3     & 1.3 & 29.1  & 3.2  & 2.4 & 4.2 \\
G358.414$-$00.122b & 358.414 & $-$0.122 & 229  & 2.6  & 0.4 & $-$213.7 & 3.5 & 50.2  & 8.6  & 2.4 & 3.5 \\
G358.460$-$00.170 & 358.460 & $-$0.170 & 229  & 4.2  & 0.4 & 10.6     & 1.0 & 24.1  & 2.5  & 2.1 & 4.5 \\
%G358.460$-$00.170b & 358.460 & $-$0.170 & 229  & 1.6  & 0.3 & $-$215.7 & 4.4 & 46.6  & 10.3 & 2.1 & 2.4 \\
%G358.506+00.250 & 358.506 & 0.250 & 229 & 7.7 & 0.9 & 24.8 & 1.0 & 18.5 & 2.4 & 7.4 \\
G358.918$-$00.195  & 358.918 & $-$0.195 & 190  & 11.1 & 0.2 & 3.9      & 0.4 & 38.4  & 0.9  & 1.5 & 20.9 \\
G003.415+00.171    & 3.415   & 0.171    & 229  & 12.7 & 0.3 & 10.3     & 0.5 & 46.6  & 1.3  & 1.7 & 23.3 \\
G004.537$-$00.288  & 4.537   & $-$0.288 & 592  & 16.2 & 0.1 & 12.5     & 0.1 & 34.1  & 0.3  & 1.2 & 36.0 \\
G007.044+00.018    & 7.044   & 0.018    & 752  & 30.1 & 0.3 & 22.8     & 0.2 & 35.0  & 0.4  & 1.2 & 67.8 \\
G009.875$-$00.160  & 9.875   & $-$0.160 & 399  & 18.2 & 0.4 & 29.7     & 0.3 & 25.5  & 0.6  & 2.1 & 20.0 \\
G012.289$-$00.327  & 12.289  & $-$0.327 & 132  & 30.9 & 0.6 & 42.9     & 0.4 & 35.5  & 0.8  & 3.6 & 23.4 \\
G012.508$-$00.127  & 12.508  & $-$0.127 & 217  & 43.0 & 0.6 & 48.8     & 0.2 & 19.0  & 0.4  & 1.8 & 47.5 \\
G012.508$-$00.127* & 12.508  & $-$0.127 & 217  & 64.7 & 1.4 & 28.7     & 0.2 & 16.7  & 0.5  & 1.8 & 67.1 \\
G012.548$-$00.293  & 12.548  & $-$0.293 & 132  & 68.1 & 1.3 & 29.7     & 0.1 & 16.2  & 0.3  & 2.4 & 52.2 \\
G012.584$-$00.119  & 12.584  & $-$0.119 & 289  & 91.3 & 1.1 & 31.1     & 0.1 & 17.3  & 0.2  & 2.0 & 86.7 \\
%G012.584$-$00.119* & 12.584  & $-$0.119 & 289  & 48.0 & 9.9 & 24.9     & 6.1 & 44.6 & 5.3  & 2.0 \\
G012.586$-$00.238  & 12.586  & $-$0.238 & 132  & 88.4 & 1.3 & 31.0     & 0.1 & 15.0  & 0.3  & 2.3 & 68.0 \\
G012.641$-$00.292  & 12.641  & $-$0.292 & 404  & 88.7 & 0.7 & 33.2     & 0.1 & 15.0  & 0.2  & 2.9 & 54.1 \\
G012.641$-$00.292* & 12.641  & $-$0.292 & 404  & 28.8 & 0.6 & 57.8     & 0.3 & 22.4  & 0.7  & 2.9 & 21.5 \\
G012.643$-$00.357  & 12.634  & $-$0.357 & 132  & 82.0 & 1.2 & 31.9     & 0.1 & 13.6  & 0.2  & 4.6 & 30.0 \\
%G012.851$-$00.175  & 12.851  & $-$0.175 & 720  & 57.6 & 0.6 & 57.3     & 0.2 & 21.4  & 0.4  & 2.3 & 52.9 \\
%G012.851$-$00.175* & 12.851  & $-$0.175 & 720  & 97.6 & 0.6 & 33.2     & 0.1 & 21.2  & 0.2  & 2.3 & 89.2 \\
G013.642+00.236    & 13.642  & 0.236    & 132  & 22.4 & 0.4 & 35.9     & 0.4 & 48.0  & 0.9  & 2.6 & 27.3 \\
G013.671$-$00.381a & 13.671  & $-$0.381 & 132  & 11.9 & 0.6 & 33.5     & 4.2 & 81.1  & 12.8 & 1.8 & 27.4 \\
G013.671$-$00.381b & 13.671  & $-$0.381 & 132  & 5.5  & 0.7 & 121.7    & 4.4 & 59.7  & 11.5 & 1.8 & 10.9 \\
G013.747$-$00.407  & 13.747  & $-$0.407 & 132  & 17.0 & 0.7 & 37.2     & 0.6 & 31.4  & 1.4  & 3.9 & 11.2 \\
G013.817$-$00.105a & 13.817  & $-$0.105 & 132  & 51.0 & 0.7 & 41.1     & 0.2 & 26.4  & 0.5  & 2.3 & 52.0 \\
G013.817$-$00.105b & 13.817  & $-$0.105 & 132  & 17.6 & 0.5 & 106.0    & 0.3 & 19.1  & 0.7  & 2.3 & 15.3 \\
G013.820$-$00.250  & 13.820  & $-$0.250 & 132  & 15.2 & 0.4 & 102.9    & 0.3 & 21.6  & 0.8  & 2.1 & 15.4 \\
G013.820$-$00.250* & 13.820  & $-$0.250 & 132  & 14.0 & 0.4 & 39.5     & 0.7 & 49.9  & 1.9  & 2.1 & 21.5 \\
G015.238$-$00.364  & 15.238  & $-$0.364 & 132  & 29.4 & 0.6 & 20.4     & 0.4 & 34.8  & 0.9  & 4.7 & 16.9 \\
G015.238$-$00.364* & 15.238  & $-$0.364 & 132  & 16.4 & 1.2 & 58.3     & 0.3 & 9.2   & 0.8  & 4.7 & 4.8 \\
G016.816+00.353    & 16.816  & 0.353    & 132  & 30.7 & 0.9 & 27.6     & 0.4 & 28.6  & 0.9  & 5.2 & 14.4 \\
G017.187$-$00.207a & 17.187  & $-$0.207 & 110  & 20.4 & 0.4 & 43.9     & 0.3 & 26.5  & 0.8  & 3.1 & 15.5 \\
G017.187$-$00.207b & 17.187  & $-$0.207 & 110  & 12.6 & 0.5 & 97.2     & 0.6 & 28.2  & 1.4  & 3.1 & 9.9 \\
G017.187$-$00.207c & 17.187  & $-$0.207 & 110  & 7.2  & 0.4 & $-$4.5   & 1.1 & 34.0  & 2.8  & 3.1 & 6.2 \\
%G017.618+00.096    & 17.618  & 0.096    & 206  & 20.4 & 0.5 & 30.9     & 0.4 & 29.3  & 0.9  & 2.6 & 19.4 \\
G017.768$-$00.075  & 17.768  & $-$0.075 & 132  & 15.0 & 0.3 & 52.0     & 0.3 & 34.3  & 0.8  & 2.1 & 19.1 \\
%G018.641$-$00.296  & 18.641  & $-$0.296 & 302  & 44.7 & 0.4 & 67.1     & 0.2 & 36.8  & 0.4  & 2.3 & 53.8 \\
%G021.668$-$00.256  & 21.668  & $-$0.256 & 482  & 11.8 & 0.1 & 67.7     & 0.3 & 46.1  & 0.6  & 1.0 & 36.6 \\
%G024.147$-$00.547 & 24.147  & $-$0.547 & 179  & 19.2 & 0.5 & 101.9 & 0.4 & 28.7 & 1.1 & 4.1 & \\
%G024.653$-$00.270  & 24.653  & $-$0.270 & 981  & 13.0 & 0.1 & 46.0     & 0.4 & 50.8  & 1.0  & 0.6 & 70.5 \\
%G024.653$-$00.270* & 24.653  & $-$0.270 & 981  & 19.8 & 0.1 & 95.3     & 0.1 & 54.1  & 0.3  & 0.6 & 110.8 \\
G025.511+00.068    & 25.511  & 0.068    & 186 & 21.3 & 0.5 & 57.2     & 0.5 & 30.0  & 1.3  & 2.2 & 24.2 \\
G025.511+00.068*   & 25.511  & 0.068    & 186 & 13.5 & 0.3 & 105.1    & 0.7 & 48.1  & 2.0  & 2.2 & 19.4 \\
G026.584+00.163    & 26.584  & 0.163    & 289  & 36.3 & 0.4 & 105.6    & 0.1 & 24.0  & 0.3  & 1.5 & 54.1 \\
G026.584+00.163*   & 26.584  & 0.163    & 289  & 7.8  & 0.3 & 32.2     & 0.5 & 27.2  & 1.4  & 1.5 & 12.4 \\
G027.128+00.010    & 27.128  & 0.010    & 425  & 39.3 & 0.4 & 92.5     & 0.2 & 28.1  & 0.4  & 2.3 & 41.4 \\
G029.388+00.156a   & 29.388  & 0.156    & 189  & 16.5 & 0.2 & 48.9     & 0.3 & 42.9  & 0.8  & 1.9 & 26.0 \\
G029.388+00.156b   & 29.388  & 0.156    & 189  & 12.9 & 0.2 & 98.3     & 0.4 & 44.5  & 1.1  & 1.9 & 20.7 \\
%G030.517$-$00.452  & 30.517  & $-$0.452 & 197  & 67.3 & 0.3 & 103.1    & 0.1 & 36.9  & 0.2  & 1.3 & 143.6 \\
G042.735$-$00.112  & 42.735  & $-$0.112 & 442  & 18.9 & 0.2 & 68.2     & 0.1 & 21.1  & 0.3  & 1.3 & 30.5 \\
G042.735$-$00.112* & 42.735  & $-$0.112 & 442  & 4.5  & 0.2 & 21.2     & 0.5 & 24.8  & 1.2  & 1.3 & 7.9 \\
G043.379$-$00.064  & 43.379  & $-$0.064 & 132  & 6.8  & 0.2 & 60.1     & 1.1 & 59.3  & 2.5  & 2.0 & 12.0

%% file: gdigs_avf2.tab
% source          & l       & b        & R   & T    & Terr& V     & Verr& FWHM & FWHMerr & rms & S/N \\
G355.612+00.314   & 355.612 & $0.314$  & 390 & 9.1  & 1.2 & 101.2 & 1.7 & 24.7 & 3.9 & 1.5 & 13.8 \\ 
G001.702$-$00.156 & 1.702   & $-0.156$ & 234 & 8.1  & 1.2 & 227.3 & 1.3 & 17.9 & 3.0 & 1.7 & 9.2 \\ 
G003.158$-$00.040 & 3.518   & $-0.040$ & 105 & 5.7  & 1.1 & 150.7 & 3.0 & 32.0 & 7.0 & 2.4 & 6.1 \\ 
G006.085$-$00.016 & 6.085   & $-0.016$ & 225 & 5.0  & 1.1 & 139.9 & 3.2 & 28.6 & 7.4 & 1.6 & 7.6 \\ 
G009.402+00.180   & 9.402   & $0.180$  & 415 & 9.2  & 1.7 & 131.5 & 0.5 & 5.1  & 1.1 & 1.4 & 6.8 \\ 
G011.936+00.020   & 11.936  & $0.020$  & 285 & 6.4  & 1.8 & 120.5 & 2.7 & 19.2 & 6.3 & 1.7 & 7.5 \\ 
G012.380$-$00.081 & 12.380  & $-0.081$ & 445 & 5.5  & 1.3 & 118.5 & 2.5 & 18.8 & 5.3 & 1.3 & 8.4 \\ 
G013.782$-$00.152 & 13.782  & $-0.152$ & 501 & 11.0 & 0.6 & 102.8 & 0.7 & 26.2 & 1.7 & 1.2 & 21.4 \\ 
G014.549$-$00.106 & 14.549  & $-0.106$ & 398 & 10.5 & 1.6 & 150.5 & 1.9 & 26.0 & 4.6 & 1.2 & 20.4 \\ 
G020.206+00.036   & 20.206  & $0.036$  & 251 & 12.6 & 1.3 & 117.2 & 0.8 & 16.3 & 1.9 & 1.9 & 12.2